\documentclass[preprint,longbibliography]{revtex4-2}
\usepackage{enumitem,hyperref}
\usepackage{xcolor}
\usepackage{amsmath,bm,srcltx}
\usepackage{amssymb,amsfonts,graphicx}%
\graphicspath{{./fig/}}
\newcommand{\de}{{\rm d}}

\newcommand{\z }{{\textstyle\frac\partial{\partial t}}}

\renewcommand{\v}{{\dot\gamma}}

\begin{document}
\title{Transient Elasticity -- A Unifying Framework for\\ Thixotropy,  Polymers, and Granular Media}
\author{Mario Liu}\email{mliu@uni-tuebingen.de}
\affiliation{Theoretische Physik, Universit\"{a}t T\"{u}bingen,\\ 72076 T\"{u}bingen, Germany, EC }
\date{\today}

%
%
\begin{abstract}
Thixotropic yield stress fluids, such as paint or ketchup, are traditionally viewed as elastic structures that, under shear rates, break into viscous liquids with 
lumps, and reconnect at rest.  An alternative framework is presented here: Upon shear, structural  destruction is incomplete, leaving sufficient connections. As a 
result, elasticity fundamentally underlies their non-Newtonian behavior, though they do appear purely viscous under a steady shear, displaying little direct evidence of elastic rebound---same as  granular media and polymers. 
Consequently, all three are described by the same set of 
evolution equations, differing only in their parameters. 

These equations are set up by starting from solid dynamics and allowing the elastic strain $
\varepsilon^e$ to relax, which in effect  interpolates between solid and fluid behavior, appropriate for systems that display both types of behavior. Incorporating in addition a two-temperature framework, tracking how energy is dissipated 
in two consecutive stages, yields the nonlinear model of \textit{Transient Elasticity} (TE). It describes an elasticity that is transient in time yet persistent under shear. 
Previously validated for polymers and granular media, TE is here applied to 
thixotropic yield-stress fluids. As shown, it successfully accounts for a wide range of 
characteristic phenomena, including over- and undershoot, viscosity bifurcation, shear banding, 
and oscillatory rheography. Given its appropriateness across structurally diverse systems, TE offers a unified, surprisingly general account of non-Newtonian phenomena.
\end{abstract}
\maketitle \newpage
\tableofcontents

\section{Introduction\label{te1}}
The energy $w$ of a Newtonian fluid depends on the state variables: densities of mass $\rho$, entropy $s$, and momentum $g_i=\rho v_i$. Their respective conjugate variables  are: the chemical potential $\mu\equiv\partial w/\partial\rho$,  temperature $T\equiv\partial w/\partial s$, and velocity $v_i\equiv\partial w/\partial g_i=\partial (g_i^2/2\rho)/\partial g_i$ (where $g_i^2/2\rho$ is the kinetic energy). 
Fluid-dynamics is given by the evolution of its state variables.
{Solid-dynamics} has an additional state variable,  the elastic strain $\varepsilon^e_{ ij}$ (the part of strain that stores the elastic energy). The elastic stress, $ {\sigma^e_{ ij}}\equiv\partial w/\partial {\varepsilon^e_{ ij}}$ is its conjugate variable. For the (1,2)-componen, the total stress is $ {\sigma}= {\sigma^e}+ \eta\v$, where  ${\v}$ is the shear rate and $ \sigma^e=K\varepsilon^e$ for linear elasticity. The coefficients $K$ and $\eta$ may depend on state variables such as $T$ and $\rho$.  For uniform stresses, the evolution for 
${\varepsilon^e}$ is the equality of both rates, $\z {\varepsilon^e}= {\v}$. 
As setting $ {\varepsilon^e}\equiv0$ reduces solid- to fluid-dynamics (see Sec.\,\ref{HoTE}),  allowing \textit{elastic decay} by letting ${\varepsilon^e}$ relax interpolates between both, 
\begin{align}\label{funda}
\z {\varepsilon^e}= {\v}- {\varepsilon^e}/\tau.
\end{align}
There is no relaxation for $\tau\to\infty$, and the system is solid. For $\tau\to0$, the 
relaxation of elasticity is instantaneous, implying ${\varepsilon^e}\equiv0$, and the system is always fluid.  This is a surgical change to solid 
dynamics:  The state variables remain the same, as does the stress, including the rate-independence of the viscosity $\eta$. Replacing $\z$ by a convective derivative, as in Sec.\,\ref{HoTE}, this model, called Transient Elasticity (TE), is as fully  nonlinear as solid dynamics. 

In a more engineering language,  Eq.(\ref{funda}) divides the rate $\v$ into an elastic and plastic part,  $\v\equiv\dot\varepsilon_ {12}=\z {\varepsilon^e}+\dot\varepsilon^p$, where $\dot\varepsilon^p= {\varepsilon^e}/\tau$   differentiates the two previously equal rates. (In contrast to the elastic strain $\varepsilon^e$, the two ``strains'':  $\gamma\equiv\varepsilon_ {12}$ and $\varepsilon^p$, are not uniquely defined, only their rates are.  Therefore, we employ different notations: $\z\varepsilon^e$ vs.\,$\v$ and $\dot\varepsilon^p$.)  

Numerous materials such as paint, mayonnaise, ketchup, are \textit{thixotropic yield-stress fluids} (TYF) that sustain an elastic structure at rest. Paint stays on the wall instead of 
dripping, ketchup stays in an open, bottom-up jar.  In the usual picture, this structure is destroyed  by shear rates or agitation,  rendering the system viscous with lumps. Hence the 
paint becomes liquid under brushing, and ketchup flows out of the jar when shaken. The lumps become smaller with increasing shear rates, decreasing the viscosity, see eg.\,
\cite{thix4a}.  This agrees with the reviews cited below and the definition of IUPAC (international union of pure and applied chemistry) for  thixotropy,  as ``\textit{the 
continuous decrease of viscosity with time when flow is applied $\cdots$ the subsequent recovery of viscosity when flow is discontinued.''}

In contrast, TE's physics describes an elastic structure that persists under a steady 
shear rate yet, surprisingly, appears viscous and liquid, if critical (ie. if stationary): For $\z {\varepsilon^e}=0$, we have $ {\varepsilon^e}={\varepsilon^c}=\tau{\v}$ from Eq.(\ref{funda}). 
As long as $\tau\not=0$,  the elastic stress is finite, yet it camouflages as a  viscous stress of a flowing medium: $\sigma^e=\sigma^c=K{\varepsilon^c}= K\tau{\v}$, apparently with the viscosity  $\sigma^c/\v=K\tau$. For nonlinear elasticity, this viscosity is rate-dependent. This elastic stress, as shown below, naturally accounts for many thixotropic phenomena. 

Usually, elasticity accounts for solids that execute reversible deformations, while viscosity accounts for liquids' irreversible motion. Yet adding one term to solid 
dynamics, as in Eq.(\ref{funda}), does render an elastic system flowing irreversibly, blurring the line between solids and liquids. (Note however, its channel of dissipation is elastic decay, $\varepsilon^e\to0$, quite different from the viscous channel that renders the velocity uniform.) 

As one typically measures the total stress,  $ {\sigma}= {\sigma^e}+ \eta\v$, not $
\sigma^e$ and $\eta\v$ separately, distinguishing such a TE-system from a viscous liquid is subtle. 
Evidence for TE, circumstantial than watertight, is provided by transient, off-critical phenomena such as oscillatory strains (considered in Sec.\ref{Gprime}). One may also adopt a bird's 
view to observe the ease thixotropic phenomena are accounted for by TE, without any additional 
construction. Finally, the fact that polymers, granular media and TYF share the same set of TE-equations can hardly be accidental, it must be the result of underlying principles of physics, as the 
three systems are microscopically vastly different.  Mainly, these are  energy conservation and proper entropy production, obeyed by solid- and fluid-dynamics, hence also by TE,  their interpolation. 
These principles are sufficiently confining to fix TE's structure, such that the differences in composition and interaction are reflected only in the parameters (such as $\tau$). 

That elasticity exists under shear may actually be seen in granular media.  
At rest, part of the grains, forming continuous \textit{force-chains} 
that  extend from end to end and carry the elastic stress load. These force chains can be rendered 
visible~\cite{fc}. Under shear rates, many of the force-chains remain. They jiggle and vibrate, jumping from grains to 
grains, but persist to exist, see the \href{https://www.youtube.com/watch?v=_gIzbiPBY8A}{video}. (Searching for \textit{``force chains''}, one finds more examples.)  
As the same physics is operative in partially saturated soils~\cite{unsat} and clays~\cite{clay1}, 
circumstances may well be similar in TYF, : As part of the structure disentangles, rips  and slips under shear, other parts take over the stress load, maintaining residual 
elasticity. The usual argument that TYF possess fragile structure that are easily destroyed because they are only loosely bound, fails to convince as grains are certainly 
looser.

For reasons given in Sec.\ref{TmDyn}, one typically has $\tau\sim1/\v$ for TYF and granular media if stationary. This is remarkable, because the elastic critical stress is then rate-independent, $\sigma^e=\sigma^c=K{\varepsilon^c}= K\tau{\v}
=$ const, implying a total stress $\sigma=\sigma^c+\eta\v$ that resembles the Bingham model. Also, assuming a purely elastic stress, $\sigma=\sigma^e$ with $\eta=0$, the Maxwell model is shown in Sec.\ref{jeffrey} equivalent to Eq.(\ref{funda}). Since Bingham and Maxwell, called respectively type 1 and type 2  model  in~\cite{thix4}, are the two common starting points to model TYF,  these relations are relevant---even though both types in effect model viscous liquids.

For all these reasons, it is useful to probe what TE 
accomplishes, to see whether it is a viable alternative. 
In  Sec.\ref{HoTE}, the TE-equations are presented, and briefly introduced. (A detailed derivation may be found in~\cite{exp31,workinput}.) The main part of the paper follows, containing calculations on rate-controlled experiments in Sec.\ref{Rate} (flow rules, rate jumps, rate ramps, oscillatory rheography),  stress-controlled experiments  in Sec.\ref{SCE} (viscosity bifurcation, stress jumps, tilted plane), and non-uniform experiments  in Sec.\ref{nue} (shear bands, non-local fluidization, elastic shear waves). All results agree well with the helpful reviews I consulted: Barnes~\cite{thix2,thix3}, Mewis and Wagner~\cite{thix1}, Larson  and  Wei~\cite{Larson}, de Souza Mendes and Thompson~\cite{thix4,thix4b,thix4c}, Mujumdar, Beris,  and Metzner~\cite{mujum}. The present work would not be possible without their well presented thoughts and results. 
Seeing that TE does this is certainly more essential than agreeing that the equations of Sec.\ref{HoTE} are physically plausible, even necessary.

Previous papers have shown that TE captures the behavior of polymers~\cite{polymer-1,polymer-1a,polymer-3,polymer-5a,polymer-5} (espedcially\cite{polymer-5a,polymer-5})  and  granular media~\cite{granL1,granL2,granL3,exp21,exp31,exp32,exp4,exp9,exp92,exp11,luding}  
 (especially~\cite{exp11}), including water retention of partially saturated soils~\cite{unsat,unsat2,workinput}. 
A more recent paper on a TE-model called \textit{Terracotta} provides a 
realistic  account of clays~\cite{clay1}. Adding TYF to them would render the list of TE-systems satisfyingly complete, and reinforce the conjecture that many non-Newtonian effects are transiently elastic in origin. 

Both fluid- and solid-dynamics comply with general principles of physics, especially energy conservation and positive entropy production, which are local versions respectively of first and second law of thermodynamics.  
Their equilibrium properties are encoded in the energy expression,     
and dynamic equations are divided into \textit{reactive} and \textit{dissipative} part---as prescribed by the Onsager relation---such that the former has zero, and the latter positive, 
entropy production.  Such models may be called  \textit{compliant}, cf.\,App.\ref{hydro}.  Any constitutive model aiming to be widely applicable needs to be compliant, as real systems in nature always are. If a 
model fails at compliance, it is bound to disagree with some observations. 
Being an interpolation of both dynamics, TE is also compliant. It has a clearly defined (solid) energy and its entropy production,  with $\eta=$ const, fully complies with the Onsager 
relation. This is the main reason TE is capable of accounting for different systems. Its structure is  determined by that of the solid-dynamics, leaving its coefficients to reflect the 
enormously varied composition and interaction.  (Though there is no  generally accepted nonlinear Onsager 
relations, some restrictions do apply, which is typically disrespected when a seemingly 
suitable $\eta(\v)$ is taken. Note $\eta$ may depend on state variables, such as temperature and density, which do not alter the entropy production, yet a dependence on $\v$ does.)


In exploring TE for how it accommodates TYF, even in circumstances that have not yet been probed experimentally, the calculations are made as simple as possible, employing  mostly linear elasticity and constant coefficients, leading to simple, analytic results. Model sculpting is generally avoided. (If more realism is desired, to fit specific experiments,  these simplifying assumptions probably need to be abandoned.) The claim that TE accounts for TYF is therefore  quick to verify, showing it is indeed a viable alternative. 


\section{The Expressions of TE\label{HoTE}}
\subsection{Elastic Relaxation}
For TYF, it suffices to take stresses, strains and rates as traceless (as is also true for polymers, but not for granular media). Hence Momentum conservation is
\begin{align}\label{eq1}
&\z (\rho v_i)-\nabla_j(\sigma_{ij}-\rho v_iv_j)=0, \,\, \sigma_{ij}=\sigma_{ij}^e+\eta \v_{ij},
\\\nonumber
&\sigma_{ij}^e\equiv\partial w/\partial\varepsilon^e_{ij},\,\, \v_{ij}\equiv(\nabla_iv_j+\nabla_jv_i)/2.
\end{align}
Assuming isotropy, $\sigma_{ij}^e=K\varepsilon^e_{ij}$ for linear elasticity. The elastic strain obeys the  equation
\begin{align}\label{eq2}
D_t\varepsilon^{e}_{ij}=\dot \varepsilon_{ij}& -  \varepsilon^{e}_{ij}/{\tau},\,\,\text{with}\,\,
\Omega_{ij}\equiv(\nabla_iv_j-\nabla_jv_i)/2  
\\\nonumber
\text{and}\,\,D_t{\varepsilon^e_{ij}}&\equiv(\z + v_k \nabla_{\! k}){ \varepsilon^{e}_{ij}}+ {\Omega_{ik}\varepsilon^{e}_{kj}-\varepsilon^{e}_{ik}\Omega_{kj}}.
\end{align}
The Einstein summation convention is employed, here and below. Note that the signs of $\v, \varepsilon^e, 
\sigma^e$ are unambiguously given by Eqs.(\ref{eq1}). 
(Coming from solid-dynamics, it appears one has to take the upper convective derivative, as in~\cite{polymer-1,polymer-1a,polymer-3,polymer-5,polymer-5a}. However, the co-rotational one is chosen for simplicity, as it is equally admissible, see~\cite{workinput} around Eqs.(18,19), and \cite{alert}.)  
For $\tau\to\infty$, Eqs.(\ref{eq1},\ref{eq2}) are the solid-dynamics; for $\tau\to0$, $\varepsilon^{e}
_{ij}={\tau}(\varepsilon_{ij}-D_t\varepsilon^{e}_{ij})\equiv0$. As the elastic energy is minimal for $\varepsilon^{e}_{ij}=0$, its derivative also vanishes, implying $\sigma^{e}_{ij}\equiv\partial w/\partial\varepsilon^e_{ij}\equiv0$. 
This reduces  Eqs.(\ref{eq1}) to fluid-dynamics. 

%
Linearizing Eqs.(\ref{eq2}) by replacing $D_t\to\z$, denoting the (1,2)-components as 
$\v_{12}\to\v$, $\varepsilon^e_{12}\to\varepsilon^e,\,\, \sigma^e_{12}\to\sigma^e, \,\, \sigma_{12}\to\sigma$,
and assuming linear elasticity, 
$\sigma^e=K\varepsilon^e$, we have
\begin{align}\label{eq3a}
\z\varepsilon^e=\v-\varepsilon^e/\tau,\,\,
\varepsilon^e=\varepsilon^{c}=\v\tau,\,\, 
\sigma^{c}=(K\tau+\eta)\v,
\end{align}
where $^{c}$ denotes the critical state, $\z\varepsilon^{c}=0$. Neglecting $\bm{\Omega\cdot\varepsilon^{e}-\varepsilon^{e}\cdot\Omega}$ of $D_t$ is possible if $\tau\v\ll1$: With $\Omega\sim\v$,  the term $\Omega\varepsilon^e\sim\v\varepsilon^e$ is then nonlinearly small. In Sec.\,\ref{critState}, the complete  $D_t$ must be included, because 
$\tau\v\sim\v/T_m^n\sim\v/|\v|^n\not\ll1$ for $\v\to0$, see Eq.(\ref{eq9a}).  

\subsection{A Second Temperature\label{TmDyn}}
A system with two temperatures is nothing unusual: Pouring cold milk into hot tea creates such a system,  briefly.  
TYF sustain two temperatures for a longer time span, call them $T$ and $T_m$, whose equilibration $T_m= T$ provides a timescale. In the literature\cite{Larson,thix1,thix2,thix3,thix4,thix4b,thix4c,mujum}, a  structure variable $\lambda$  
is employed that does the same. So they serve the same purpose mathematically, though physically there are a few reasons  to prefer $T_m$: First, it is a general quantity, usefull also in granular media, and it is in principle measurable. Second, setting up a 
\textit{compliant} theory, both temperatures have to be included  
to account for proper energy dissipation. Third, the dynamics of a temperature is severely constrained by thermodynamics, rendering $T_m$ less flexible \textit{a priori}, though more convincing if it works,  \textit{a posteriori}.

Generally speaking, temperature is the average energy of a group of \textit{degrees of freedom}. 
If there are two groups, their respective temperatures may easily be different. 
There are many such systems, a well known one is a fully ionized plasma~\cite{plasma}:  Ions and electrons collide quasi-elastically, transferring momentum while  keeping  their respective energy, such that a plasma has one macroscopic velocity but two temperatures. 
Granular media also have two temperatures: The usual temperature $T$ quantifies the energy of atoms, the granular temperature $T_g$  characterizes    
the random motion and deformation of the grains~\cite{athermal,plasma}. 
Water and air in soil~\cite{unsat,unsat2}, and Clays~\cite{clay1},  have part of their atoms organized in mesoscopic structures. The elastic and kinetic energy of their vibration is quantified by the temperature ${T_m}$ (for \textit{mesoscopic temperature}). Equilibration is typically slow. 
The same circumstance should also apply to TYF: $T$ for the atoms and $T_m$ for the flocs and clusters. If the flocs are strongly elongated, as with polymer solutions, equilibration  $T_m\to T$ is, due to the large surface to bulk ratio, sufficiently fast that $T_m\approx T$ is not an independent state variable.

The dynamics for a second temperatures 
is always the same, determined by energy conservation, entropy production, and  \textit{two-stage irreversibility}: The energy that goes into $T_m$ from viscous heating is withdrawn from the macroscopic kinetic energy; the ensuing relaxation of $T_m$ feeds the energy into $T$, into atoms. As dictated by \textit{statistical mechanics}, the direction of feeding is always towards the smaller, much more numerous group:
{\bf macro $\to$ meso $\to$ micro},  
never backwards.\cite{athermal}

The entropy $s$ and meso-entropy $s_m$ are defined as the logarithm of the number of  atomic and mesoscopic degrees of freedom, respectively, with $T\equiv\partial w/\partial s$ and 
$T_m\equiv\partial w/\partial s_m$, hence $\de w=T\de s+T_m\de s_m$ Rewriting it as 
$\de w=T\de(s+s_m)+(T_m-T)\de s_m$, we have an energetic minimum in equilibrium, at $T_m=T$,  for given total entropy $s+s_m$, or $T_m-T=\partial w/\partial s_m|_{s+s_m}=0$. Expanding the energy  $w(s+s_m,s_m)$ around a given total entropy $s+s_m$, we have 
$w= s_m^2/2A+\cdots$ with $A>0$ and $T_m-T=s_m/A$.
The comparative massiveness of meso-structures, hence large kinetic energy, implies $T_m\gg T$, or $T_m-T\approx T_m$. Therefore,  equilibrium may 
be approximated as $T_m=0$, instead of $T_m=T$. This is similar to a pendulum that, due to its thermal energy, oscillates slightly in equilibrium, yet is well approximated as being at rest, with zero  motion. The equation for $s_m$ is then
\begin{align}\label{eq6}
\z s_m+&\nabla_i(s_mv_i-\kappa\nabla_i T_m)=(R_m-\beta T_m^2)/T_m,
\\\nonumber&R_m=\kappa(\nabla_iT_m)^2+\eta\v_{ij}\v_{ij}>0, 
\end{align}
where $\kappa,\eta,\beta>0$. 
The terms $s_mv_i$ is  convective, $\kappa\nabla_i T_m$  diffusive. 
The meso-entropy production $R_m$ increases $s_m$ and accounts for the dissipation into the first of the \textit{two-stage irreversibility}---mainly by  viscous heating $\sim\v^2$.
The term $-\beta T_m$ relaxes $s_m$, until $T_m=0$.  
Except for  $-\beta T_m$, these are the same expressions of any normal,  one-temperature system, fluid or solid. The detailed derivation of Eqs.(\ref{eq6}), from energy conservation, entropy production, and two-stage irreversibility, is 
given in  Sec.\,(7.1) of~\cite{exp31}, see also App.\ref{dd}. 

Assuming uniformity, $\nabla_iT_m,\nabla_iv_i=0$, Eqs.(\ref{eq6}) reduce to
\begin{align}\label{eq8}
&\z T_m^2=-r_m(T_m^2- f^2\v^2),\,\,\text{implying}
\\\label{eq9}
&(T_m/f)^2=\v ^2+(\v_i^2-\v ^2)e^{-r_mt},
\end{align}
where $r_m\equiv2\beta/A$, $ f^2\equiv\eta/\beta$. Eq.(\ref{eq9}) is the solution after a rate jump at $t=0$, from $\v_i$ to $\v$. 

For $\z s_m=A\z T_m=0$ and weak non-uniformities, $(\nabla_iT_m)^2\ll T_m\nabla_i^2T_m$, we have 
\begin{align}\label{eq8a}
\nabla_i^2T_m^2=(T_m^2-f^2\v^2)2\beta/\kappa,
\end{align}
which is  relevant for shear bands, at the fluid-solid interface, cf.\,Sec.\,\ref{nf}.  

We take $f=1$, because it determines how much $\v$ raises $T_m$, in units of $T$. Yet because we approximate  $T_m=T$ with $T_m=0$, we never compare 
$T_m$ to $T$, and are free to measure $T_m$ in the unit of $\v$. 
Note that changing $f$ does not change its dynamics, see Eq.(\ref{eq9}) or (\ref{eq8a}). For uniform, stationary states, we therefore have 
\begin{align}\label{Tm dot gamma}
T_m=|\dot\gamma|.
\end{align}

Because thermal excitations jiggle flocs randomly and loosen the structural entanglement, they enable and facilitate elastic decay, hence $1/\tau$ should grow with $T_m$, and we take  
\begin{align} \label{eq9a}
1/\tau=\lambda_n T_m^n,\, n>0\,(\text{mostly}\,\,n=1).
\end{align}
This is a constitutive choice. 
As stated above, TE's structure is largely determined by those of the solid-dynamics, leaving its coefficients to reflect the enormously varied composition and interaction of the three TE-systems. 
Unfortunately, the connection between coefficients and microscopics  is usually subtle, as we  well know from fluid-dynamics: Its structure holds for many different fluids, but the values of viscosity and heat diffusion coefficient, especially  their functional dependences on  density and temperature, are hard to calculate and usually deduced from experiments. Similarly, Eq.(\ref{eq9a}) is shown  appropriate for all experiments considered below, including  the flow rules of the next section, see Fig.\ref{flowrules}. Should some experiments contradict it, a better choice needs to be found. 

For granular media, $n=1$ alone is appropriate; for polymers, $n=0$ is. As explained above, this is due to the large surface to bulk ratio. All  three systems share the same dynamical structure---though of course their composition and interactions are very different.

\subsection{Bingham and Maxwell Models vs. TE\label{jeffrey}}
To convince the reader of TE's worthiness, the main trust of this paper is algebraic, not conceptual.  
For the present paper, the purpose is to show---in Sec.\ref{Rate}, \ref{SCE}, \ref{nue}---that the equations of TE are well capable of accounting for  a wide range of thixotropic effects.
TE is given by Eqs.(\ref{eq1},\ref{eq2},\ref{eq6}) and the unchanged rest of solid dynamics. It has two 
relaxation times, $\tau$ for elasticity,  $1/r_m$ 
for $T_m$, and the length scale $\sqrt{\kappa/2\beta}$, characterizing the diffusion of $T_m$. These three  parameters are in addition to those already contained in solid-dynamics. Both $r_m$ and the viscosity $\eta$ are taken as constant in this paper. Only $\tau$ depends on $T_m$. 

This section is an exception. It embeds TE with two familiar models, Maxwell and Bingham. And by clarifying that viscous dissipation and elastic relaxation differ qualitatively, it finds Maxwell to be a purely elastic model.  
As concepts such as \textit{time inversion parity} (that is 
basic to Onsager relations) is employed below, some results may appear sudden for a person unfamiliar with it. They may simply skip such discussions. 

Taking thixotropy as a viscous property, it is typically modeled   along two seemingly orthogonal  lines, 
either by starting from \textit{viscoplastic}, Bingham-like expressions, or  
from \textit{viscoelastic}, Maxwell/Jeffrey-like expressions, see the above cited reviews, especially the careful analysis in~\cite{thix4}. 
This section shows that although linearized TE is equivalent to a specific Jeffrey model, it is also well capable of producing Bingham behavior. 

The Bingham model is a postulated  stress-vs-rate curve, given by the steady flow state, 
$\sigma=\sigma^{\rm dY}+\eta\v$, with a const $\sigma^{\rm dY}$ called \textit{dynamic yield stress}.  This behavior is in fact contained in TE:   
Viscous heating by shear flow is a general phenomenon, leading to $T_m=|\v|$ in steady state, see Eq.\ref{Tm dot gamma}. 
If Eq.(\ref{funda}) is also stationary, with $1/\tau\sim T_m$, Eq.(\ref{eq9a}), the  {critical} elastic strain $\varepsilon^c$ is rate-independent: 
$\varepsilon^e=\varepsilon^c=\v \tau\sim\v/T_m\sim\v/|\v|=\pm1$. As a result,  the stress $
\sigma=\sigma^c+\eta\v$ acquires the Bingham form, where $\sigma^{\rm dY}
\equiv\sigma^c=K\varepsilon^c$ for linear elasticity. The system flows at any given rate $\v$,  but only for $\sigma>\sigma^{c}$ at given stress. It comes to rest for  $
\sigma<\sigma^{c}$, as observed, and named \textit{viscosity bifurcation}~\cite{vicBifurc,vicBifurc2}, see Sec.\,\ref{vibi}. 

\begin{figure}[t]
\begin{center}
\includegraphics[scale=.5]{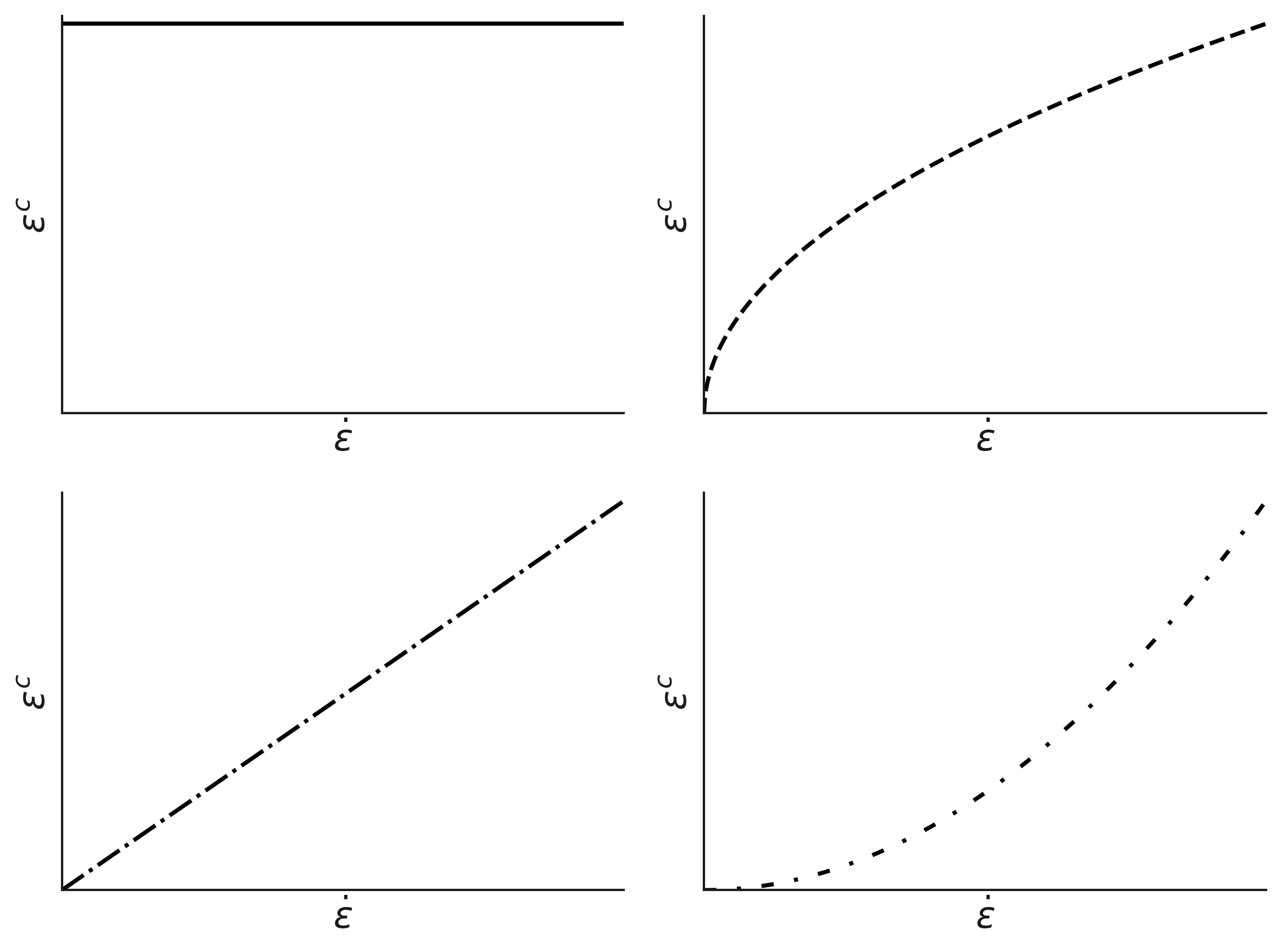}\caption{\label{flowrules}
\textbf{flow rules}:  Steady-state elastic strain $\varepsilon^c$ vs.\,shear rate $\v$, assuming $1/\tau\sim T_m^{n}$, with the elastic stress $\sigma^c=K\varepsilon^c$. For low rates $\v>0$ and $n=1, 0.5, 2, 3$, TE yields  respectively: $\varepsilon^c=$ const [Bingham model], $\varepsilon^c\sim\sqrt{\v}$ [shear thinning], $\varepsilon^c\sim\v$  [Newtonian], and  $\varepsilon^c\sim\v^2$ [shear thickening]. } 
\end{center}
\end{figure}

More generally, we take: $1/\tau\sim T_m^n\sim|\v|^n$, $n>0$ from Eq.(\ref{eq9a}). As shown in 
Sec.\,\ref{critState}, this leads (including the convective derivative and at sufficiently low rates) for $n= 0.5, 2, 3$ respectively,  to $\sigma^e\sim\sqrt{\v}$ (shear thinning), $
\sigma^e\sim\v$ (Newtonian), and  $\sigma^e\sim\v^2$ (shear thickening),  see Fig.\,\ref{flowrules}. Clearly,  these typical non-Newtonian effects do not need to be viscous. 

Moreover, different exponents may be combined, say $1/\tau\sim T_m\sim|\v|$ for low rates, and $1/\tau\sim T_m^3\sim|\v|^3$ for higher ones, 
which yields Herschel–Bulkley behavior. Combining $1/\tau\sim T_m^2\sim|\v|^2$ for low rates,  and   $1/\tau\sim T_m\sim|\v|$ for higher ones leads to Fig.\,\ref{fig1}. 

Next, we show that linearized TE is equivalent to a specific Jeffrey model:  
Taking $\bar\eta\hat=K\tau+\eta$ and $\bar\lambda\hat=\eta\tau$ in the Jeffrey model, $\sigma+\tau\z\sigma=\bar\eta\v+\bar\lambda\z\v$, we arrive at 
$\z(\sigma-\eta\v)=K\v-(\sigma-\eta\v)/\tau$, 
same as Eq.(\ref{funda}) if $\sigma=K\varepsilon^e+\eta\v$ holds. In other words, the special Jeffrey expression, $\sigma+\tau\z\sigma=(K\tau+\eta)\v+\eta\tau\z\v$,
with $K,\eta=$const and $\tau$ given by Eq.(\ref{eq9a}) may equally well be employed to calculate all linear  effects obtained in this paper (even though it is mathematically more complex). 
For  $\eta=0$, implying $\sigma=K\varepsilon^e$, we have the Maxwell model with a purely elastic stress. 

This may be a bit surprising, because Maxwell is usually deduced by combining $\v=\sigma/\bar\eta$ for fluid and $\v=\z\sigma/K$ for solid, yielding $\v=\sigma/\bar\eta+\z\sigma/K$ 
that identifies $\bar\eta\v$ as the viscous stress, naming $\bar\eta$ the Maxwell viscosity. Yet  this interpretation  contradicts Eq.(\ref{funda}), a relaxation equation 
that confines the duration of elasticity (hence the name TE), because elastic relaxation  and viscous damping are two distinct channels of dissipation. The former diminishes elasticity, $
\varepsilon^e\to0$, and the latter let the velocity become 
uniform (or a solid-rotation), $\v\to0$. The first dissipates elastic, the second kinetic, energy, with the respective rates given by $(K/\tau)(\varepsilon^e)^2$ and 
$\eta\v^2$, see Eq.(\ref{2-18a}).  
In Eq.(\ref{funda}), $\z {\varepsilon^e}$ is odd under {time inversion}. Being also  odd, $\v$ is---according to the Onsager Relation~\cite{OnsagerLIT}---reactive, reversible, and does not produce entropy, while ${\varepsilon^e}/\tau$ is even, irreversible, and produces entropy at the rate $(K/\tau)(\varepsilon^e)^2$. 
The two corresponding terms in the Maxwell model, $\v=\sigma/\bar\eta+\z\sigma/K$, are $\v$ and $\sigma/\bar\eta$, where the former is reactive, and the latter dissipative. Not producing any entropy, $\bar\eta\v$ is not the viscous stress, and $\bar\eta$ not a viscosity.  
\textit{Maxwell viscosity} is a misnomer.

Requiring Maxwell/Jeffrey to reduce to Eq.(\ref{funda}) renders them compliant. Arbitrary coefficients do not do this. 
In terms of mechanical analogues, the  Maxwell model is given 
by a spring $K$ and a dashpot $\bar\eta$ in series, while Jeffrey and TE are given by the Maxwell strand in parallel with a dashpot $\eta$, see eg.\,Fig.\,2 of \cite{thix4}. Clearly, a dashpot simply accounts for the spring\rq{}s damping. It does not distinguish  between elastic relaxation ($\bar\eta$) and viscosity ($\eta$).

\section{Rate-Controlled Experiments\label{Rate}}
Moving one plate at constant speed or constant force while holding the other implies, respectively,  rate- or stress-control. This leads to significant differences.  All calculations in this section assume rate-control, with $\v$  given. We take 
$1/\tau=$ {const} and $1/\tau_n=\lambda_nT_m^n$ {with} $n>0$, see Eqs.(\ref{eq9a}).
Since $\tau(T_m)=\tau(|\v|)=$ const, both share the same critical state. 

\subsection{The Critical  States \label{critState}}
Denoting $\varepsilon^e\equiv\varepsilon^e_{12}=\varepsilon^e_{21}$, $\varepsilon^d\equiv\varepsilon^e_{11}=-\varepsilon^e_{22}$, and taking the velocity as ${v}_i = [2\v y, 0]$, we have, from  Eqs.\,(\ref{eq2}),
\begin{align*}
&{\v}_{ij}=
\begin{pmatrix}
0 & \v \\
\v & 0
\end{pmatrix}\!\!,\,\,\,
{\Omega}_{ij}=
\begin{pmatrix}
0 & -\v \\
\v & 0 \quad
\end{pmatrix}\!\!,
\\
&{\Omega_{ik}\varepsilon^{e}_{kj}-\varepsilon^{e}_{ik}\Omega_{kj}}
={2\v}\cdot\!
 \begin{pmatrix}
-\varepsilon^e & \varepsilon^d  \\\nonumber
\varepsilon^d  & \varepsilon^e 
\end{pmatrix}\!\!,
\end{align*}
with which we rewrite the first of   
Eqs.\,(\ref{eq2}) (assuming $\nabla_{k} \varepsilon^{e}=0$) as
 \begin{align}\label{eq12}
\z  &\varepsilon^d(t) = 2\v\,\varepsilon^e - \varepsilon^d/\tau,\quad \z  \varepsilon^e(t) = \v(1 -2\varepsilon^d) - \varepsilon^e/\tau; 
\\\nonumber
&\text{or for}\,\, z\equiv\varepsilon^e+i\varepsilon^d: \,\,\,\,\,  \z  z(t)=\v -(1/\tau-2i\v )z.
\end{align} 
We call the double stationary solution,  $T_m=|\v|$  and
$\z z=0$ ($\z\varepsilon^e =0$, $\z\varepsilon^d=0$) \textit{critical},  denoting it with the superscript $^c$. We have $z^c=\v\tau/(1-i2\v\tau)$, or 
\begin{align}
\label{eq5}
\varepsilon^{c}_{ij}&=\frac{\v\tau}{1+(2\v\tau)^2}
\begin{pmatrix}
2\v\tau & 1\\
1 & -2\v\tau 
\end{pmatrix}\\\nonumber
&=\frac{1}{4+(\v\tau)^{-2}}
\begin{pmatrix}
2 & 1/\v\tau\\
1/\v\tau & -2 
\end{pmatrix}.
\end{align}
Inserting $1/\tau=\lambda_n T_m^n=\lambda_n|\v|^n$, 
with $\pm\equiv\frac\v{|\v|}$, $X_n\equiv\pm\lambda_n|\v|^{n-1}$, $\varepsilon^{c}\equiv\varepsilon^{e}_{12}$, we have
\begin{align}\label{eq10}
\varepsilon^{c}_{ij}(n)=&
\frac{1}{4+X_n^2}
\begin{pmatrix}
2 & X_n\\
 X_n & -2 
\end{pmatrix}\!\!, \quad\text{implying}
\\\nonumber
\varepsilon^{c}(0.5)&=\frac{\pm\lambda_{0.5}\sqrt{|\v|}}{4+\lambda_{0.5}^2|\v|},\quad\varepsilon^{c}(1)=\frac{\pm\lambda_1}{4+\lambda_1^2},
\\\nonumber
\varepsilon^{c}(2)=&\frac{\lambda_2\v}{4+(\lambda_2\v)^2}, \quad \varepsilon^{c}(3)=\frac{\pm\lambda_3\v^2}{4+(\lambda_3\v^2)^2},
\end{align}
see Fig.\,\ref{flowrules}. As long as $\v$ keeps its sign, $\varepsilon^{c}(1)$ is rate-independent, while $\varepsilon^{c}(2)$ fakes a  Newtonian viscosity for 
$\lambda_2\v\ll1$, showing shear thinning otherwise. 
These results depend on $D_t$, as the low-rate regime  $X_n\ll1$ actually implies $\v\tau=1/X_n\gg1$.

The critical state $\varepsilon^{c}(1)$ is surprisingly appropriate for TYF: 
It remains constant, however small $\v$ is. And since the stress $\sigma=K\varepsilon^{c}(1)+\eta\v$ takes on the Bingham behavior, with $\sigma/\v$ diverging for $\v\to0$, one may take $\sigma^c=K\varepsilon^{c}$ as the  \textit{dynamic yield-stress}.  For  $n\not=1$, we see from Eqs.(\ref{eq10}) that  
 \begin{align}
\varepsilon^{c}(n)\stackrel{\v\to0}{\to}0,\quad \varepsilon^{c}(n)\stackrel{\v\to\infty}{\to}0. 
\end{align}
The first relation shows that $n=1$ is the only choice truly compatible with \textit{viscosity bifurcation}, see the considerations of Fig.\,\ref{fig5}. The second implies $\sigma=K\varepsilon^{c}+\eta\v$ reduces to $\eta\v$ for $\v\to\infty$. Note $\eta$, the actual viscosity,  is frequently denoted as $\eta_\infty$. (For $n=1$, we also have $K\varepsilon^{c}(1)\ll\eta\v$ for sufficiently high rates.)

\begin{figure}[t]
\begin{center}
\includegraphics[scale=.5]{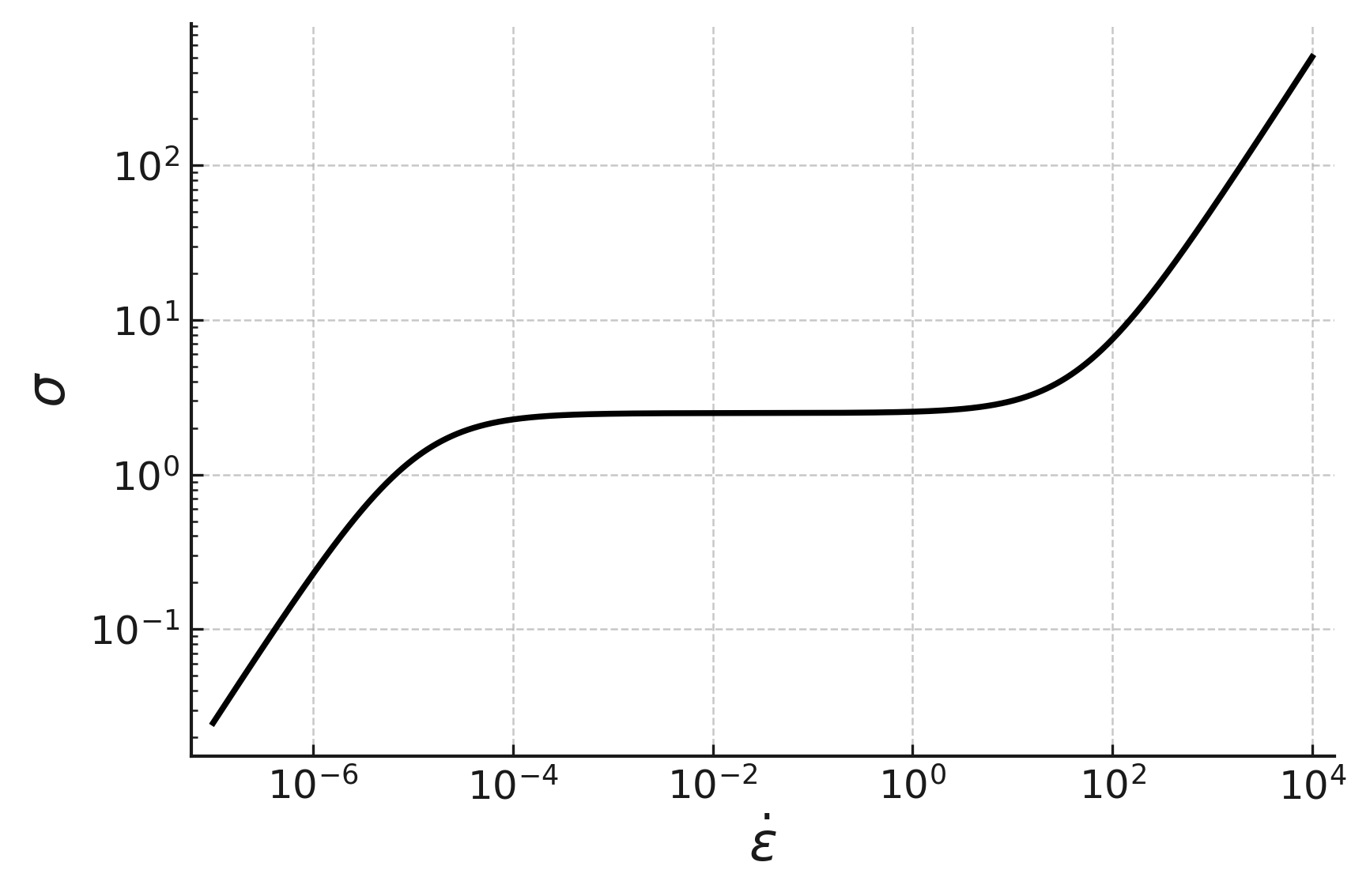}
\caption{\label{fig1}A log-log plot of the critical stress $\sigma=K\varepsilon^c+\eta\v $ vs.\,rate $\v$, with $\varepsilon^c$ of Eq.(\ref{eq5}), $\tau$ of Eq.(\ref{eq11}), and the parameters in the text. It is essentially the same plot as Fig.\,6 of the \textit{``everything flows''}-review 
by Barnes~\cite{thix3}. There are two (Newtonian) viscosity plateaus, for $\v>10^3$ and $\v<10^{-5}$, which sandwich a long stretch of constant stress $K\varepsilon^c$, for 6 orders of magnitude of $\v$.} 
\end{center}
\end{figure}

\textbf{Hybrid Flow Rules}\quad The critical state $\varepsilon^{c}(1)$ may not be what is actually realized in nature, as it has problems with 
equilibrium fluctuations of $T_m$. As will be shown in Sec.\,\ref{vibi}, $T_m$ relaxes to zero for $
\v\equiv0$, only if the equilibrium stress is smaller than the dynamic yield stress, $\sigma<\sigma^c$; it grows  for  $\sigma>\sigma^c$. As  $T_m$-fluctuations are always present, stress values $\sigma^c<\sigma<\sigma^{\rm Y}$ are not stable---contrary to observation, say on a tilted surface, see Sec.\,\ref{tiltedP}. (Stresses above $\sigma^{\rm Y}$ are never stable, see Sec.\,\ref{jeffrey}, \ref{yield}.) With  $\varepsilon^{c}(2)$ and $1/\tau_2=\lambda_2 T_m^2$, this instability does not exist.  Therefore, the  realized critical state may be a hybrid one that  turns  $\varepsilon^{c}(1)$ into  $\varepsilon^{c}(2)$ for $T_m\to0$, say
\begin{align}\label{eq11}
\frac1\tau=\frac{\lambda_1 T_m^2}{T_0+T_m},\quad T_0=\frac{\lambda_1}{\lambda_2}.
\end{align}
It reduces to $\lambda_1 T_m$ for $T_m\gg T_0$, and to $
\lambda_2 T_m^2$ for  $T_m\ll T_0$. 
Plotting the critical stress 
$\sigma=K\varepsilon^c+\eta\v$ using this $\tau$, with $\varepsilon^c$ 
from Eq.(\ref{eq5}), and the parameters 
\begin{quote}
$K=10^7$Pa, $\eta=0.05$Pa$\cdot$s, $\varepsilon^c_1=\lambda_1/4=10^{-6}$, $\varepsilon^c_2/\v=\lambda_2/4=10^{-1}$s,\\ implying $\lambda_1, \lambda_2\v\ll1$, $K\varepsilon^c_1=10$Pa, $K\varepsilon^c_2/\varepsilon\hat=10^6$Pa$\cdot$s, $\eta=0.05$Pa$\cdot$s
\end{quote}
we find a  [$\sigma$ vs $\v$]-curve, Fig.\ref{fig1}, that is remarkably similar to Fig.\,6 of the well-known \textit{everything flows}-review by Barnes~\cite{thix3}. 

This hybrid $\tau$ has further consequences that are relevant.
Because  Eq.(\ref{eq2}) reads $D_t\varepsilon^{e}=  \v -  \varepsilon^{e}\lambda_1|\v|$, such that  both right-hand terms are of the same order in $\v$. It remains a TE-system however small $\v$ is, and never 
becomes a genuine solid. With $1/\tau_2=\lambda_2 T_m^2$, on the other hand,  Eq.(\ref{eq2}) reads $\z\varepsilon^{e}=  \v -  \varepsilon^{e}
\lambda_2|\v|^2$, such that the second term, quadratically small,  may be neglected for $\v\to0$. This would render the system a  true solid, in which eg.\,elastic shear waves (of  weak-amplitudes) propagates with no constraints, cf.\,Sec.\,\ref{waves}. 

There are two ways to obtain a Herschel-Bulkley form. In the first, one takes $1/\tau=\lambda_1T_m[1+(T_m/T_0)^n]$, $T_0^n=\lambda_1/\lambda_{n+1}$, with $\varepsilon^{c}$ given by Eq.(\ref{eq10}). It reduces to $
\lambda_1T_m$ for $T_m\ll T_0$, and to $\lambda_{n+1}T_m^{n+1}$ for $T_m\gg T_0$. Or one takes $\sigma=K\varepsilon^{c}(1)+\eta(T_m)\v$, with 
\begin{align}\label{eq11b}
\eta(T_m)=\eta_0 T_m^n=\eta_0|\v|^n.
\end{align}
Three final points:  $\bullet$ The normal stress difference, for $1/\tau=\lambda_1T_m$ and linear elasticity, is 
rate-independent: $N_1\equiv\sigma_{11}-\sigma_{22}=K/[1+(2\lambda_1)^{-2}]$.
 $\bullet$ We may consider more velocity fields,  say elongational flow, $\bm{v}=(\dot\epsilon x, -\dot\epsilon y)$, with $\Omega_{ij}=0$, $\v_{11}=-\v_{22}=\dot\epsilon$, see Eqs.(\ref{eq1},\ref{eq2}),  yielding the critical state $\epsilon^c\equiv\varepsilon^c_{22}=-\varepsilon^c_{11}
=\tau\dot\epsilon$. With $1/\tau_n=\lambda_nT_m^n\to \lambda_n\dot\epsilon^n$, cf.\,Eq.(\ref{eq9a}), implying a constant elastic strain, $\epsilon^c_1$ for $n=1$, and $\epsilon^c_2\sim\dot\epsilon$ for $n=2$.   $\bullet$ In the following, for simplicity,  we frequently write $\varepsilon^{c}$ for $\varepsilon^{c}(1)$.

%


\subsection{Rate-Jumps \label{jumps}}
After a rate-jump, $\v_i\to\v $, we consider how the system goes from the initial critical state, $\varepsilon_i^c=\tau\v_i$ (assuming first  $\v\tau\ll1$) to the final one, $\varepsilon_f^c=\tau\v$.  
Before the jump, we have stationarity, $\z\varepsilon^e=\v_i-\varepsilon^e/\tau=\v_i-\varepsilon^e\lambda_1|\v_i|=0$. Right after it, this is no longer the case, 
$\z\varepsilon^e=\v -\varepsilon^e\lambda_1|\v_i|\not=0$, because it takes time for $T_m$ to acquire the new stationary value, going from $\v_i$ to $\v$. 
If the jump  is up, $\v >\v_i$, we have $\z\varepsilon^e>0$, hence $\varepsilon^e$ overshoots, before returning gradually to $\varepsilon^c(1)$. If the jump is down, $\v<\v_i$, we have $\z\varepsilon^e<0$, hence $\varepsilon^e$ undershoots. The same physics holds including the convective derivative, and for $1/\tau=\lambda_nT_m^n$.

Rate ramps, where $\v$ changes continuously from $\v_i$ to $\v_f$, operates under the same logic: $T_m$ lags behind, is always too small ramping up, causing $\varepsilon^e$ to hover above the instantaneous critical value. Ramping down, $T_m$ is too large, causing $\varepsilon^e$ to stay below it. The hysteresis vanishes if the $\v$-change is sufficiently slow.

\subsubsection{Polymeric Jumps}\label{pj}
\begin{figure}[t]
\begin{center}
\includegraphics[scale=.5]{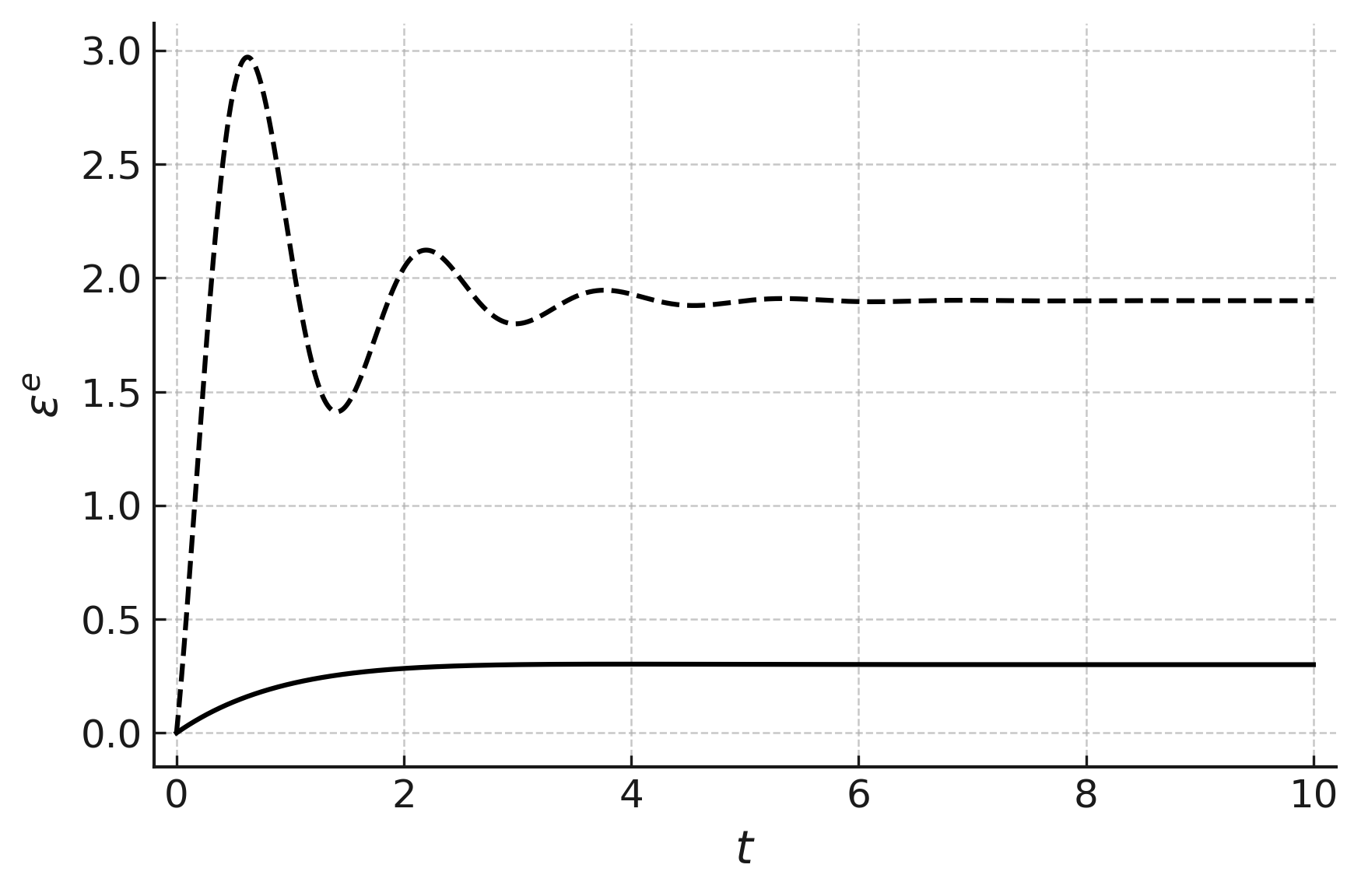}
\caption{\label{fig2}Polymeric  rate-jump behavior, with $\tau=1$s, given by elastic strain $\varepsilon^e(t)$ vs.\,time $t$, after the rate jump at $t=0$, from $\v_i$ to $\v_f $. It is drawn employing Eqs.(\ref{eq14}), with $\v_i\tau=0.1$, where the continuous line (zoomed up by a factor of 3 for a clearer view) has  $\v_f \tau=0.2$, and the dashed line has $\v_f \tau=2$, hence  showing the oscillation produced by the convective derivative. If damping were stronger, leaving only a quarter oscillation, the curve would resemble overshoot. The ordinate, starting at zero, shows the jump-initiated change in  $\varepsilon^e(t)$.}  
\end{center}
\end{figure}

In the following, we  first calculate the rate-jump behavior for a constant $\tau$ and  $\v\tau\ll1$. 
The result is a monotonic strain relaxation,  from $\varepsilon_i^c=\v_i\tau$ to $\varepsilon_f^c=\v\tau$. Next,  including the convective derivative $D_t$ to allow $\v\tau\not\ll1$, we 
find in addition a rotation of $\varepsilon^e\equiv\varepsilon^e_{12}$ vs.\,$\varepsilon^d\equiv\varepsilon^e_{11}$ that stems from $\Omega$ of $D_t$. If the damping is stronger (smaller $\tau$), and only the first quarter period of it is seen, it may easily be be confused with thixotropic overshoot, see Fig.\,\ref{fig2}. 

For $\v\tau\ll1$, Eqs.(\ref{eq3a}) holds, with ${\varepsilon^d\equiv0}$. The solution (solid line of Fig.\,\ref{fig2}) is
\begin{align}\label{eq13}
\varepsilon^e=\varepsilon^c_f+(\varepsilon^c_i-\varepsilon^c_f)e^{-t/\tau}, \,\,\varepsilon^c_i=\v_i\tau,\,\,\varepsilon^c_f=\v \tau.
\end{align}
For  $\v\tau\not\ll1$, we solve  Eqs.(\ref{eq12}). Denoting the initial and final critical states as $z_i={\v_i\tau}/({1-i2\v_i\tau})$, $z_f={\v \tau}/({1-i2\v \tau})$ [cf.\,Eqs.(\ref{eq5})], we 
add a homogeneous solution $\sim e^{-(1/\tau-i2 \v )t}$ to $z_f$ to yield the solution: 
$z=z_f+[z_i-z_f]e^{-(1/\tau-i2\v )t}$, or
\begin{align}\label{eq14} 
\varepsilon^e(t)=&\varepsilon^c_f+ [\Delta\varepsilon^c \cos(2\v  t) - \Delta\varepsilon^d \sin(2\v   t)]e^{-t/\tau} ,
\\\nonumber
\varepsilon^d(t)=&\varepsilon^d_f+ [{\Delta\varepsilon^d\cos(2\v  t)} + {\Delta\varepsilon^c\sin(2\v  t)}]e^{- t/\tau},
\end{align}
where $\Delta\varepsilon^c\equiv\varepsilon^c_i-\varepsilon^c_f$, $\Delta\varepsilon^d\equiv\varepsilon^d_i-
\varepsilon^d_f$. It combines an exponential decay with a rotation of $\Delta\varepsilon^c$ versus $
\Delta\varepsilon^d$, see Fig.\,\ref{fig2}.  (The non-monotonicity is caused by the second timescale $1/\v$.) If $\tau$ were smaller and decay faster, such that the oscillation lasts only a quarter period, the curve would resemble thixotropic overshoot.

\subsubsection{Thixotropic jumps\label{overshoot}}

Next we consider rate-jumps assuming $1/\tau=\lambda_1T_m$, for three regimes: $\tau\gg 1/r_m$, $\tau\ll 1/r_m$, and $\tau$ comparable to $1/r_m$. 
For  $\tau\gg 1/r_m$, as  $T_m=|\v |$ is  (on the timescale of the experiment)
arrived at instantly, only the relaxation of $\varepsilon^e_i\to\varepsilon^e_f$ is observed,  same as in Sec.\,\ref{pj}, there is no overshoot. 
Yet because  $\varepsilon^e_i=\varepsilon^e_f$ now, there is no response at all to the rate jump---though the stress  $\sigma=K\varepsilon^c+\eta\v$ does change via $\eta\v$. However, response is restored if the rate changes its sign, cf.\,Eqs.(\ref{eq10}). Hence rate-inversion $\v_f=-\v_i$ is considered below. (This case is appropriate for granular media, and possibly also for  {``non-thixotropic''} yield-stress fluids. Both are solids for $\v\to0$ yet do not overshoot.) 

\begin{figure}[t]
\begin{center}
\includegraphics[scale=0.5]{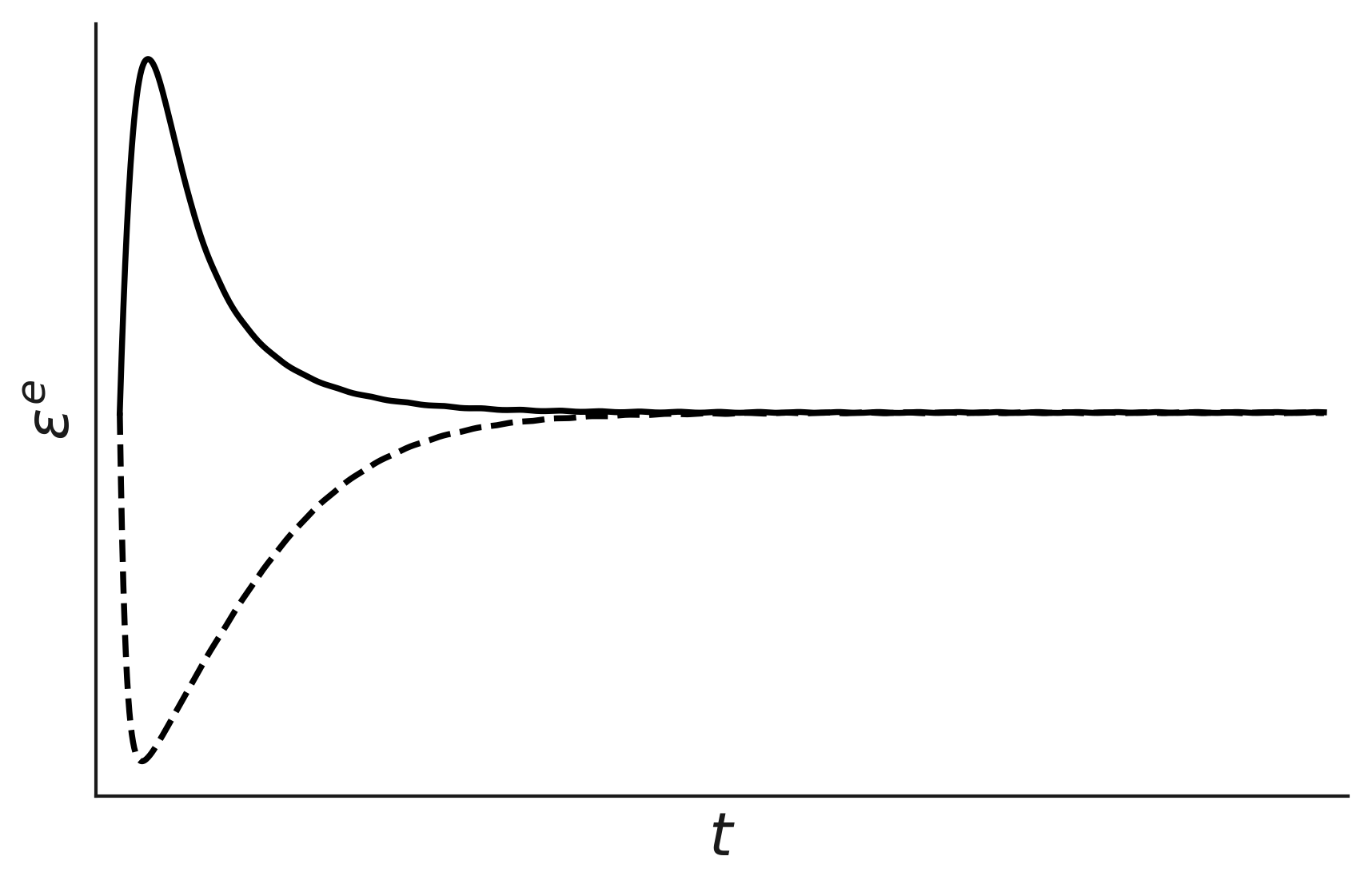}
\caption{\label{fig4}Elastic strain $\varepsilon^e(t)$ vs.\,time $t$, after a rate jump 
$\v_i\to\v $ at $t=0$, in a thixotropic fluid with $1/\tau=\lambda_1T_m$, 
showing over- and undershoot. The curves are obtained by inserting Eq.(\ref{eq9})   
and $1/\tau=\lambda_1T_m$ into Eqs.(\ref{eq3a}), 
taking:\, $\v =2$/s, $r_m=3$/s, $\lambda_1=10$.  The initial conditions 
are $\varepsilon^e_i=1/\lambda_1$, and $T_m=\v_i=1$/s  for the upper curve,  $T_m=\v_i=4$/s for the lower one. (Overshoot depends on $\tau,1/r_m$ being comparable. In this case, $\tau=1/(\lambda_1\v)=1/20$ s  and $1/r_m=1/3$ s.)
}
\end{center}
\end{figure}

For $1/r_m\gg \tau$, the $\varepsilon^e$-curve is discontinuous at $t=0$, followed by a monotonic 
approach to $\varepsilon^e_f$. It is now the elastic relaxation that happens instantly, such that Eqs.\,\,(\ref{eq5}) always hold, with $1/\tau=\lambda_1T_m\not=\lambda_1\v$. Assuming $1/\lambda_1\ll1$, the elastic strain is 
$\varepsilon^e=\v\tau=\v/(\lambda_1T_m)$, with $T_m$ given by Eq.\,\,(\ref{eq9}). Hence  $\varepsilon^e_i=1/\lambda_1$ for $t<0$,  jumping to $\v_f/(\v_i\lambda_1)$ at $t=0$ 
then returning gradually to $1/\lambda_1$. This behavior,  termed ``ideal thixotropic,'' is well illustrated by Fig.1(c) of both~\cite{Larson,thix1}. 
(Similarly,  Fig.1(b) illustrates the polymeric jump, and  Fig.1(d) the case of 
proper over- and undershoot, observed when $\tau$ and $1/r_m$ do not differ by orders of magnitude.) Finally, inserting Eq.(\ref{eq9}) and $1/\tau=\lambda_1T_m$ (implying $\varepsilon_i^c=\varepsilon_f^c$) into  Eqs.(\ref{eq3a}), taking  $1/\lambda_1=0.1$ (implying $1/r_m$ and $\tau$ are comparable), Fig.\,\ref{fig4} is numerically plotted. 

\subsubsection{Rate Ramps}
As the rate ramps up and down,  $\z\v\not=0$, different stress behavior for $\z\v>0$ and $\z\dot \varepsilon<0$ is observed. This is the continuous version of rate jumps. Ramping up fast enough, $T_m$ and $1/\tau$ are too small, rendering $\varepsilon^e$ above the instantaneous critical value. Conversely, $T_m, 1/\tau$ are too large going down, and $\varepsilon^e$ stays below. The curve going down is always below that going up. The higher the $\z\v$-rate, the more pronounced the effect. At sufficiently slow rates, there is no hysteresis.

\subsubsection{Rate Inversions \label{inversion}}
In their review article, Larson and Wei discussed rate inversion in the context of \textit{kinematic hardening} and the  \textit{Bauschinger effect}. Hence this is also considered here.
A rate-inversion is a special jump, $\v\to-\v$. We take  $1/\tau=\lambda_1T_m$, assuming first the simpler case that $T_m=|\v|$ is instantly realized (appropriate for \textit{non-thixotropic yield stress fluid}) and $\v\tau=1/\lambda_1\ll1$,  Eqs.(\ref{eq13}). The jump leads to time-dependence, because $\varepsilon^c_i=\pm1/\lambda_1$ and $
\varepsilon^c_f=\mp1/\lambda_1$ are different.  We consider three jumps, taking $t/\tau=\lambda_1T_mt=\lambda_1|\v|t$ in Eqs.(\ref{eq13}), or

\begin{center}
\begin{tabular}{||c c c c c||} 
 \hline
rate jumps & $t$ & $\varepsilon^c_i$ & $\varepsilon^c_f$ &  time dependence\\
 \hline \hline 
 $0\to\v$ & $0\to1$ & 0 &$1/\lambda_1$ & $e^{-\lambda_1|\v|t}$\\
\hline
 $\v\to-\v$ & $1\to 3$ & $1/\lambda_1$ & $-1/\lambda_1$ & $e^{-\lambda_1|\v|(t-1)}$\\
\hline
 $-\v\to\v$ & $3\to5$ & $-1/\lambda_1$ & $1/\lambda_1$ & $e^{-\lambda_1|\v|(t-3)}$\\
\hline \end{tabular}
\end{center}
\begin{figure}[h]
\begin{center}
\includegraphics[scale=0.4]{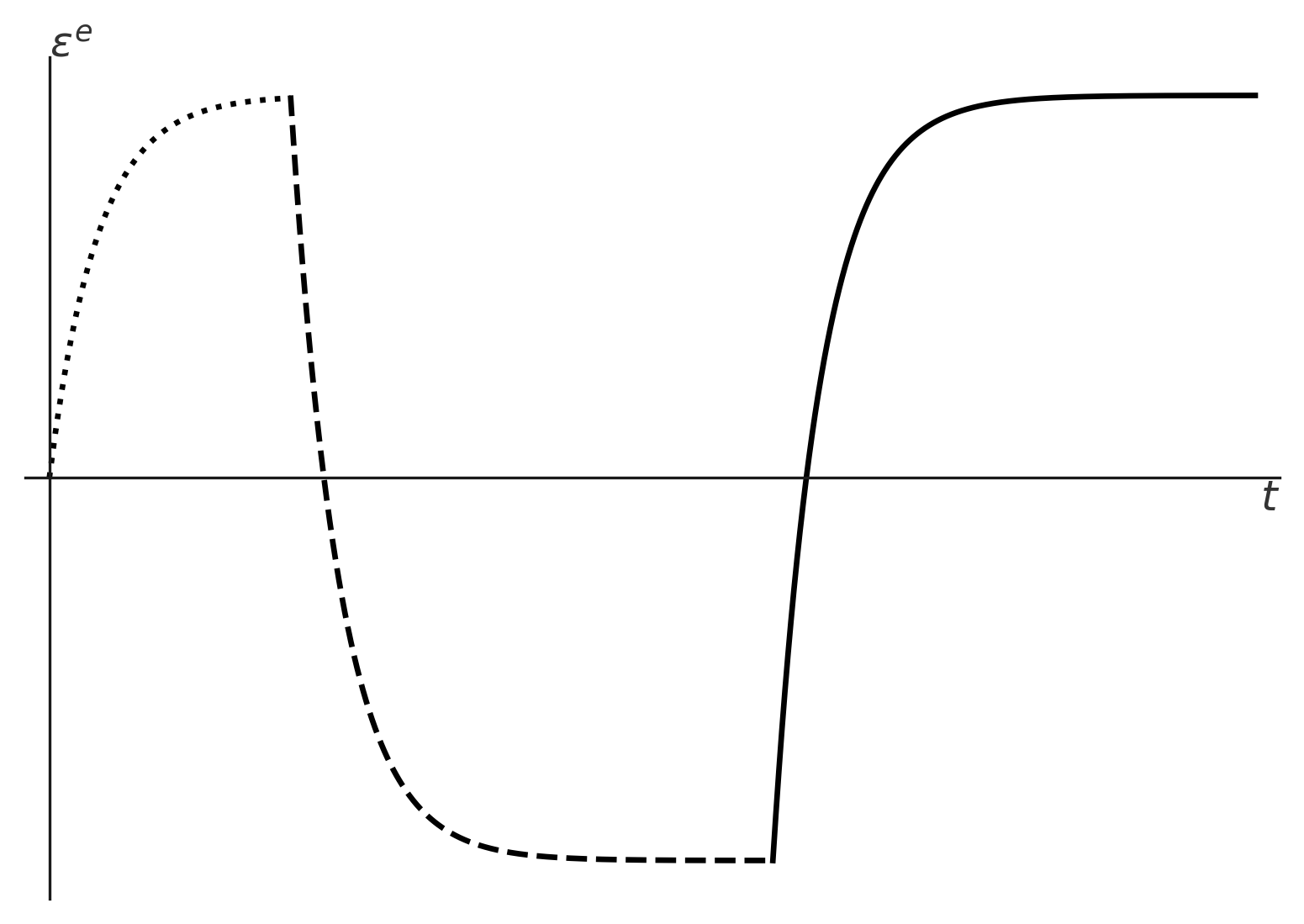}\qquad
\includegraphics[scale=0.49]{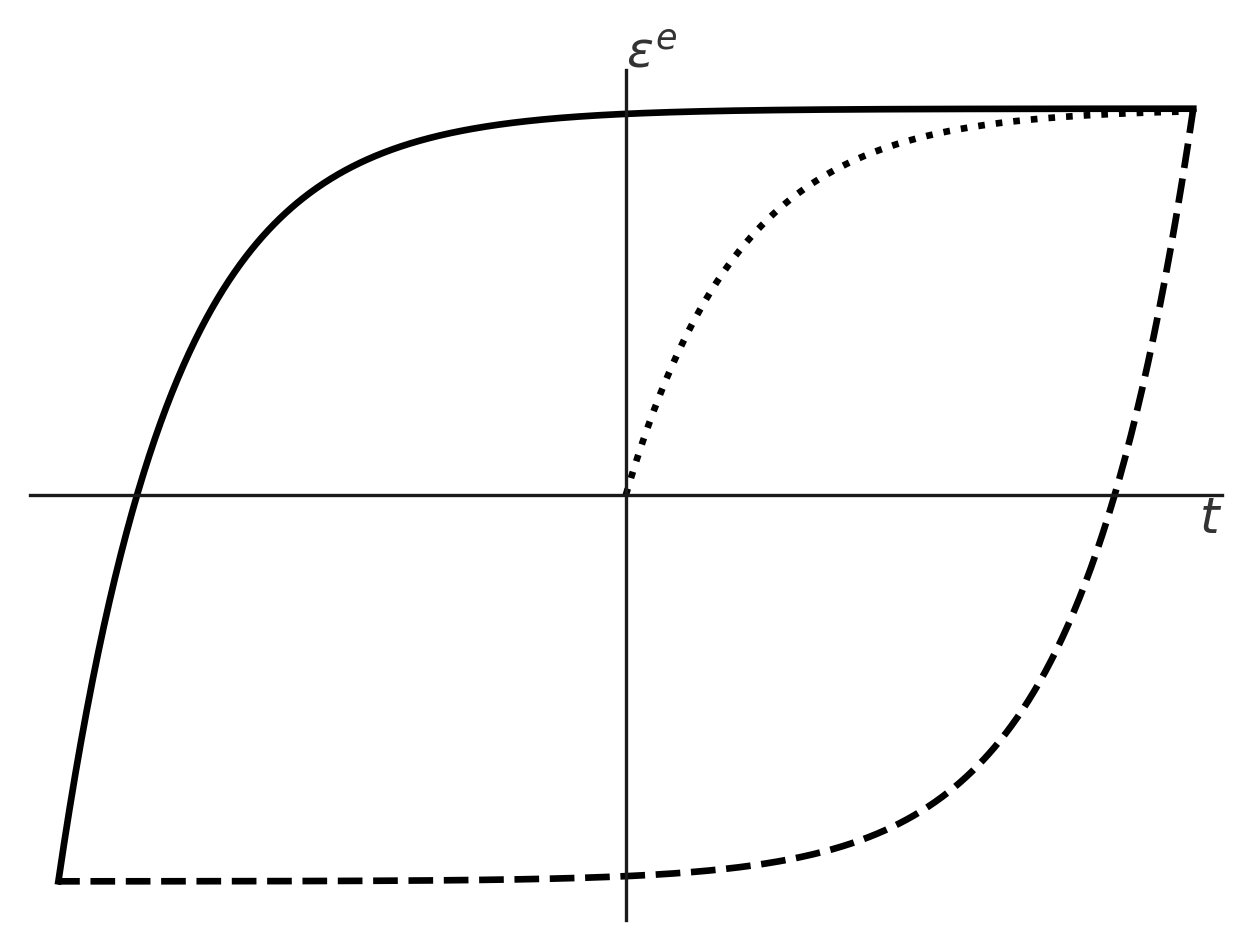}
\caption{\label{fig3}The first figure is a plot of the evolution of the elastic strain after three consecutive jumps, dotted line: 
$0\to\v$, dashed line: $\v\to-\v$, and solid line: $-\v\to\v$,   with $\lambda_1=10, \v=0.5$/s. The second figure (with the time $t$ stretched by $\sim$2x) turns the time backwards for the dash line, to suggest a dependence of the elastic strain $\varepsilon^e$ (or stress)
on the total strain $\varepsilon=|\v| t$, thus causing hysteresis.
}
\end{center}
\end{figure}
The result is the left plot of Fig.\,\ref{fig3}, showing a simple relaxation behavior after the jumps. If $T_m=|\v|$ is not 
instantly realized, as in any proper thixotropic systems, $T_m$ has to be slowly raised from 0 to $|\v|$ after the first jump, leading to overshoot, same as in Sec.\,\ref{overshoot}. Whether the second  jump leads to undershoot, depends on how quickly 
the jump takes place, how much time $T_m$ has, to relax back to zero.  If $\v$ changes to $-\v$ quickly, without stopping at $\v=0$ sufficiently long for $T_m$ to relax appreciably, $T_m=|\pm\v|$ remains unchanged, neither does the stress evolution, and there is no undershoot.  Same  holds for the third jump. 
This experiment, typically referred to as \textit{load-unload},  is popular in granular media and clays, less so in thixotropic liquids, though it could of interests here, too. 

Because of a perceived  analogy to solid, 
one is tempted to plot the stress as a function of the total strain $\varepsilon\equiv\v t$. In truth, however, the time dependence comes in the form of $|\v|(t-t_0)$, with $t_0$ the initial time. Defining  $
\varepsilon\equiv|\v| (t-t_0)$ implies that negative values of  $\varepsilon$ are given for  $t<t_0$, and $\varepsilon$ decreasing means $t$ going backwards. Handled this way, the stress appears to depend on $
\varepsilon$ alone, independent of the time.  However, plotting 
$\varepsilon^e$ (or $\sigma^e=K\varepsilon^e$) against $\varepsilon$ leads to hysteresis. (In an ideal crystal, this does not happen, because $\varepsilon=\varepsilon^e$.) The right of 
Fig.\,\ref{fig3} (with larger horizontal scale) is such a plot, with  Eqs.(\ref{eq13}) and 
\begin{center}
\begin{tabular}{||c c c c c||} 
 \hline
rate jumps & $t$ & $\varepsilon^c_i$&$\varepsilon^c_f$ &  time dependence\\
 \hline \hline 
 $0\to\v$ & $0\to 1$ & $0$&$1/\lambda_1$ & $e^{-\lambda_1|\v|t}$\\
\hline
 $\v\to-\v$ & $1\to -1$ & $1/\lambda_1$&$-1/\lambda_1$ & $e^{\lambda_1|\v|(t-1)}$\\
\hline
 $-\v\to\v$ & $-1\to1$ & $-1/\lambda_1$&$1/\lambda_1$ & $e^{-\lambda_1|\v|(t+1)}$\\
\hline \end{tabular}
\end{center}

\subsection{Oscillatory Rates: $G^\prime, G^{\prime\prime}$\label{Gprime}}
Given an oscillatory rate: $\delta\v\sim e^{-i\omega t}$, the elastic strain $\delta\varepsilon^e$ 
also oscillates,   as $\z\delta\varepsilon^e=\delta\v-\delta\varepsilon^e/\tau$,  Eq.(\ref{eq3a}), or $\delta\varepsilon^e/\delta\gamma=-i\omega \tau/(1-i\omega\tau)$, where  $
\delta\gamma\equiv\delta\v/(-i\omega)$ is the strain amplitude. Hence for $G\equiv G^\prime-iG^{\prime\prime}\equiv\delta\sigma/\delta\gamma=(K\delta\varepsilon^e+\eta\delta\v)/\delta\gamma$, we 
have 
\begin{align}\label{eq50}
{G^\prime}=\frac{K\omega^2\tau^2}{1+\omega^2\tau^2},\,\,
{G^{\prime\prime}}=\frac{K\omega\tau}{1+\omega^2\tau^2}+\eta\omega.
\end{align}
If $\tau=$ constant, these are the familiar expressions of the Jeffrey model. Sweeping $\omega$, one has $G^\prime$ going from 0 to $K$, and $G^{\prime\prime}$  maximal at  $\omega\tau=1$. 

\begin{figure}[t]
\begin{center}
\includegraphics[scale=0.6]{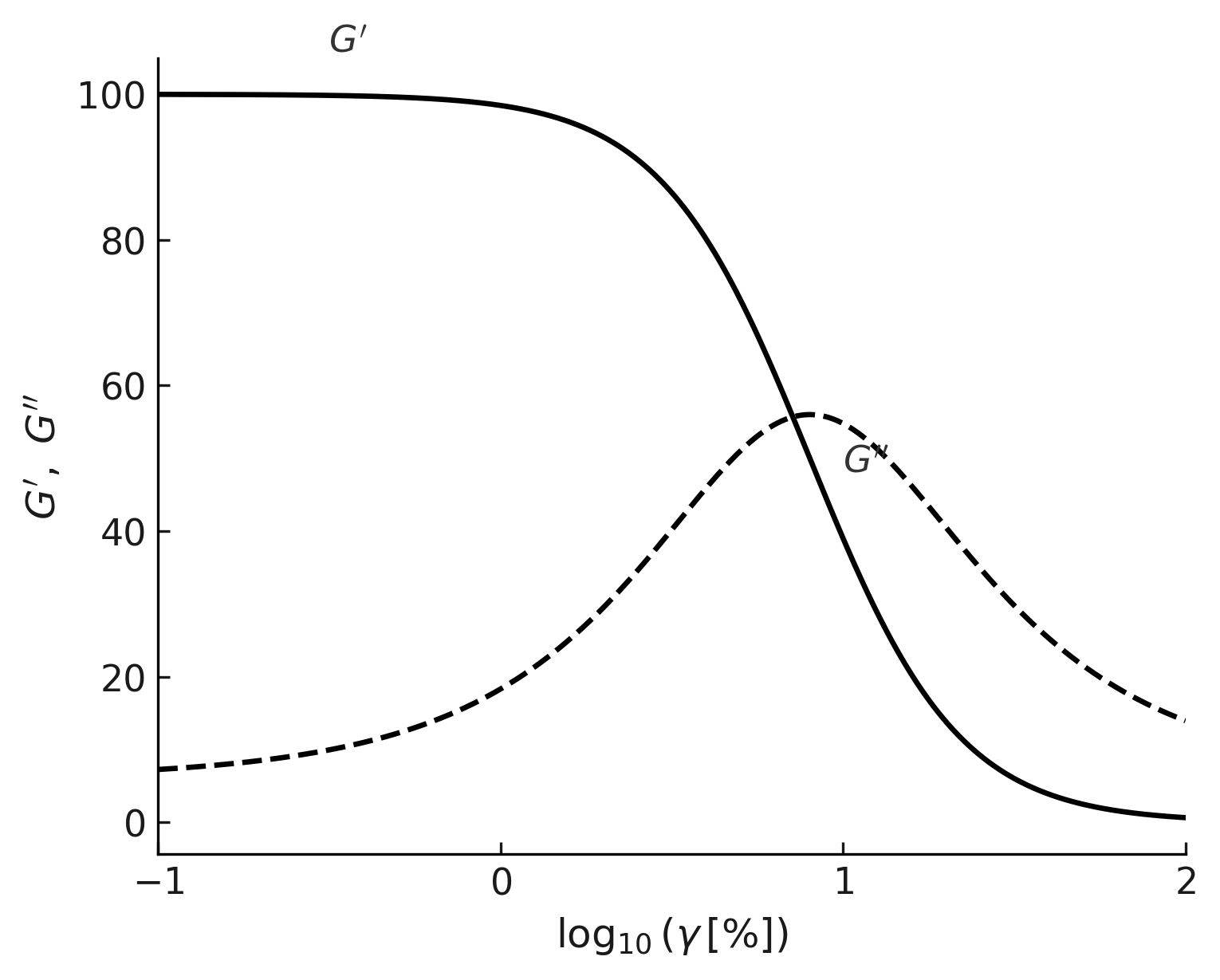}
\caption{\label{figX}A plot of $G^{\prime}$ (solid) and $G^{\prime\prime}$ (dashed) vs.\,the oscillation amplitude, $\log_{10}
(\gamma[\%])\equiv\log_{10}(100\bar\gamma/2\pi)$, as given by Eqs.\,(\ref{eq60}), taking  
$K=100$ Pa,  $\eta\omega=6$ Pa, and $\lambda_1=2$. It is quite close to the measured  Fig.\,2 
of~\cite{aging} (whose $G^{\prime\prime}$ is maximal at $\gamma=8\%$, or $\bar\gamma=2\pi 8/100\approx0.5$, implying $\lambda_1=2$).}
\end{center}
\end{figure}

For $1/\tau=\lambda_1T_m$, circumstances change drastically: For $\omega\ll r_m$, since $T_m$ follows $\v$ instantly, the term $
\delta\varepsilon^e/\tau=\lambda_1\delta\v\delta\varepsilon^e$ is quadratic,  leading to anharmonicity. 
For  $\omega\gg r_m$, $T_m$ is too slow to follow, it grows slowly until the time average of  Eq.(\ref{eq8}) vanishes,   $T_m^2=\langle\delta\v^2\rangle=\omega^2\langle{\gamma}^2\rangle$, or 
\begin{align}\label{EQ60}
T_m/\omega=\langle|
\gamma|\rangle\equiv\bar\gamma,
\end{align}
implying $\omega\tau=\omega/\lambda_1T_m=1/(\lambda_1\bar{\gamma})$, and
\begin{align}\label{eq60}
{G^\prime}=\frac{K}{1+(\lambda_1\bar\gamma)^2},\,\,
{G^{\prime\prime}}=\frac{K\lambda_1\bar\gamma}{1+(\lambda_1\bar\gamma)^2}+\eta\omega, 
\end{align}
as plotted in Fig.\,\ref{figX}. 
It agrees well with the experiment  by  C.\,Derec et al,  
see Fig.\,2 of \cite{aging} (in which $G^{\prime}, G^{\prime\prime}$ were given on a logarithmic scale, with $G^{\prime}=K=10^{100}$~Pa, $G^{\prime\prime}=\eta\omega=10^6$~Pa 
for $\gamma[\%]\to0$.  As these appear implausibly  large, $K=100$~Pa and $\eta\omega=6$~Pa were taken instead, noting that the original case can also be easily fitted). It is serendipitous, as one does not expect  a  linear calculation to account  for amplitude-dependence. 
With $\omega\tau=1/(\lambda_1\bar\gamma)$, the dependence is backwards: 
Increasing $\omega$ is the same as decreasing $\bar\gamma$. All else remains the same, especially that $G^{\prime\prime}$ is maximal at $\omega\tau= 1/\lambda_1\bar\gamma=1$. For $\bar\gamma$ large, we have $G^{\prime}\to K/
{\lambda_1^2\bar\gamma^2}$ vanishing quadratically, faster than the linear $G^{\prime\prime}-\eta\omega\to K/{(\lambda_1\bar\gamma)}$. 

Plotting $G^{\prime}, G^{\prime\prime}$ vs.\,$\omega$ for given $\bar\gamma$, the agreement is just as good. Although Eqs.\,(\ref{eq60}) show $G^{\prime}, G^{\prime\prime}-\eta\omega=$ constant, while  Fig.\,3 of~\cite{aging} shows  $G^{\prime}, G^{\prime\prime}-\eta\sqrt\omega$ is approximately constant, the disagreement is easily ameliorated by taking $n=-0.5$ in Eq.(\ref{eq11b}: The viscous stress is then $(\eta_0/\sqrt{T_m})\v=(\eta_0/\sqrt{\omega\bar\gamma})\omega\bar\gamma\sim\sqrt\omega$.

Considering this experiment under a steady shear  $\v_0$, with $\sigma_0=(K\tau+\eta)\v_0$, and taking $\sigma=\sigma_0+\delta\sigma$, we see that the quotient $\delta\sigma/\delta\varepsilon$ is still given by Eqs.(\ref{eq50}), though now with  $T_m^2=\langle(\v_0+\delta\v)^2\rangle=\v_0^2(1+\omega\bar\gamma/\v_0)^2$, or for $\omega\bar\gamma/\v_0\ll1$,
\begin{align}\label{eq61}
T_m=|\v_0|.
\end{align}
This restores the viscoelastic dependence of $G^\prime, G^{\prime\prime}$ on $\omega$, and eliminates that on $\bar\gamma$. 
Verifying these would  show the basic point unambiguously, that elasticity exists post-yield, that force-chains are not completely destroyed.  


\subsection{Aging and Rejuvenation\label{aging}}

As observed by many, see the review articles of Sec.\,\ref{te1},  the stress overshoot after a rate jump increases with the waiting time $t_w$ at rest -- called \textit{aging and rejuvenation}. In the case of  C.\,Derec et al, $t_w$ ranges from $10^2$  to $10^4$ s, see Fig.\,8 of~\cite{aging}. This is a different timescale than $\tau$ or $1/r_m$. Yet introducing a new state variable with such a timescale appears to be overdoing the modeling, as this alters the  structure of TE, risking phantom predictions.  
In other words, this effect does not ``pre-exit'' in TE---unlike those considered elsewhere, especially the Bingham-like critical state, the oscillatory coefficients: $G^\prime, G^{\prime\prime}$, and viscosity bifurcation of Sec.\,\ref{vibi}. Some model sculpting  is necessary here. The TE modifications below are tentative, qualitative, serving merely to point out a possible path.

The maxima of  overshoots for different $t_w$ coincide, see Fig.\,8 of~\cite{aging}. This indicates we may leave the dynamics of $\varepsilon^e$ unchanged, 
and employ an elastic coefficient $K$ that becomes larger for $T_m\to0$. This is also  physically plausible, as the structure does get stronger with less 
excitation. (As stated in~\cite{aging}, \textit{the experiments...are extremely sensitive to noise, particularly the ones with a waiting time.}) We take, with $\alpha, \Delta, T_1=$ const, 
\begin{align}\label{eq40}
K=K_0[1+\Delta \exp{(-\alpha T_m^2/T_1^2)}],
\end{align}
such that $K=K_0(1+\Delta)$ for $\alpha T_m^2\ll T_1^2$, and $K=K_0$ for $\alpha T_m^2\gg T_1^2$.  Only when the waiting time $t_w$ is sufficiently long, does the stress overshoot with $\Delta\not=0$. 

Furthermore, we need to slow down the exponential relaxation of $T_m$ when it is very small, by replacing $r_m$, say,  with $r_m/(1+T_1^2/T_m^2)$, changing Eq.(\ref{eq8}) to
\begin{align}\label{eq41}
\z T^2_m=-r_m\frac{T_m^2-\v^2}{1+T_1^2/T_m^2},
\end{align}
such that little changes for $T_m^2\gg T_1^2$, 
yet for $T_m^2\ll T_1^2$,  the dynamics becomes $\z T^2_m=-(r_m/{T_1^2}){T_m^2}(T_m^2-\v^2)$, reflecting  a slow, algebraic decay, whose solution, for $\v=0$ and the initial value $T_{i}$ at $t=0$, is the slow, inversely linear evolution, ${T_m^{2}}/{T_{i}^{2}}=[1+{r_m}({T_i^{2}/{T_{1}^{2}}})t]^{-1}$. Inserting  Eq.(\ref{eq41}) into (\ref{eq40}), and noting that $T_m$ does not change quickly enough,  shows an aging effect that, with appropriate values for $\alpha, \Delta, T_1$, should be close to observation. 


\section{Stress-Controlled Experiments\label{SCE}} 
\subsection{\label{yield}  Static Yield-Stress ${\sigma^{\rm Y}}$}  
Solid systems cannot sustain arbitrarily high stresses at rest. There is always  a yield-stress $\sigma^{\rm Y}$, at which the system looses elasticity, starts to move, is no longer in equilibrium. Such \textit{static yield} is the actual  plastic transition, not the Bingham-like critical state. Since static yield starts  in a quiescent system, for $\v=0$, its value is encoded in the energy $w$ (see Sec.\,\ref{te1} on compliant theories). Only the post-yield behavior is given by the  dynamics. 

The static yield stress, $\sigma^{\rm Y}\equiv\sigma^e(\varepsilon^{\rm Y})$ occurs at $\v=0$, an 
example is a lump of fluid at rest on a tilted plane that starts to flow when the tilt reaches a 
certain angle. It is different from the dynamic yield stress  $\sigma^{c}$, the rate-independent  critical state, see Eq.(\ref{eq3a}).  That a single yield stress is conceptually inappropriate was noted by  
Larson and  Wei~\cite{Larson}, who stated that yield is \textit{``the minimum stress required to trigger a 
major microstructure breakdown in an initially fully-structured material. It is thus a quantity related to the fully structured state only. Hence, to use the term yield stress to denote a quantity that is a function of the 
structuring level is clearly not consistent with the concept of yield stress''.} 
In the same vein, Mewis and Wagner~\cite{thix1} wrote:  \textit{``It could be argued on physical grounds that attributing a Bingham yield stress to a structure that exists during flow is not suitable.''}

Thermodynamic stability with respect to fluctuations of the state variables requires the energy $w$ to be a convex function of them. This is why compressibility and specific heat, both second 
derivatives of $w$, are positive. This also applies to elasticity: If the elastic part of $w$ is convex with respect to the elastic strain, the system is stable towards elastic fluctuations.  Yield occurs at the inflection 
point of $w$, where it turns concave, eg. 
\begin{align}\label{eq19}
w=w_0\left( s, \rho \right)+\textstyle\frac12K_1\varepsilon^{e}_{ij}\varepsilon^{e}_{ij}  -\textstyle\frac14K_2\varepsilon^{e}_{ij}\varepsilon^{e}_{jk}\varepsilon^{e}_{k\ell}\varepsilon^{e}_{\ell i}.
\end{align}
Since the energy is minimal for $\varepsilon^{e}_{ij}=0$, $K_1$ must be positive. 
Given any state variable $z$, the energy $w$ is convex  for $\partial^2 w/\partial z^2>0$. 
More generally, writing $ \varepsilon^{e}_{ij}$ as a 6-tuple vector, $\varepsilon^{e}_{\alpha}$ with $\alpha=1,2,\dots, 6$, {stability} is given if its Hessian matrix
has only positive eigenvalues. Otherwise, there is no elasticity. 
In our case, with $x=\varepsilon^e_{12}=\varepsilon^e_{21}$, $y=\varepsilon^e_{11}=-\varepsilon^e_{22}$,  the Hessian matrix is 
\begin{align}\nonumber\begin{pmatrix}
\partial^2 w/\partial x^2&\partial^2 w/\partial x\partial y\\
\partial^2 w/\partial y\partial x&\partial^2 w/\partial y^2
\end{pmatrix}.\end{align}
With 
$\varepsilon^{e}_{ij}\varepsilon^{e}_{ij}=2(y^2+x^2)$,
$\varepsilon^{e}_{ij}\varepsilon^{e}_{jk}\varepsilon^{e}_{k\ell}\varepsilon^{e}_{\ell i}=2(y^2+x^2)^2$,
we have
$w=w_0+K_1(y^2+x^2)-\textstyle\frac12K_2(y^2+x^2)^2$, 
with the stress and Eigenvalues given as
\begin{align}\nonumber
\sigma^{e}_{ij}=[K_1-K_2(y^2+x^2)]\varepsilon^e_{ij},\\
\lambda_1= 2K_1-6K_2 (y^2+x^2),\\\nonumber
\lambda_2=2K_1-2K_2(y^2+x^2).
\end{align}
We have $\lambda_1,\lambda_2>0$ for $K_2<0$, with an infinite elastic region. For  $K_2>0$,  the region is confined by the yield circle (similar to \textit{Hencky-von Mises} condition), 
\begin{align}\label{eq18}
x^2+y^2=(\varepsilon^e)^2+(\varepsilon^{d})^2< K_1/3K_2.
\end{align}
A drop of TYF at rest, sitting on a tilted plane of angle $\alpha$, has $\varepsilon^e\sim\sin\alpha$, and (neglecting the weight of the drop) $\varepsilon^d=0$. It will start moving if this condition is breached.  
In a shear flow, we have $y^2+x^2={(\v\tau)^2}/[1+(2\v\tau)^2]$, cf.\,Eqs.(\ref{eq5}). As Eq.(\ref{eq18}) also applies to it, elastic shear flow must obey the inequality,  
\begin{align}
(\v\tau)^2<{K_1}/({3K_2-4K_1}).
\end{align}


\subsection{Viscosity Bifurcation\label{vibi}}

\begin{figure}[t]
\begin{center}
\includegraphics[scale=.5]{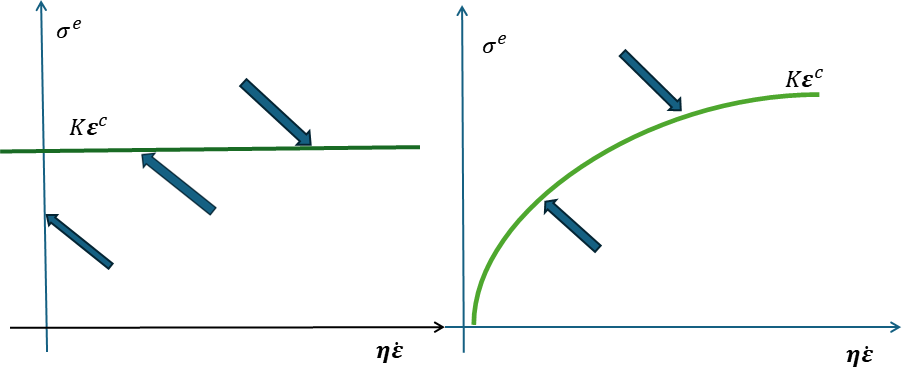}
\caption{\label{fig5}\textit{Viscosity bifurcation} as described by Eq.(\ref{eq51},\ref{eq52}): In the left figure, with $1/\tau=\lambda_1 T_m$,  the elastic critical stress $K\varepsilon^c$ is a horizontal line. The three arrows show the start and end of $T_m,\v$-relaxations, 
with their angles fixed to $\pi/2$ by $\z\sigma=K\z\varepsilon^e+\eta\z\v =0$. The first arrow from left 
shows $\sigma=K\varepsilon^e+\eta\v<K\varepsilon^c$, with $r_T>0$, hence $T_m\sim\v$ 
relaxes, with $\sigma^e$ growing, arriving at $\sigma^e=\sigma$ and $T_m,
\v=0$. The second arrow has  $\sigma>K\varepsilon^c>K\varepsilon^e$ initially, still $r_T>0$. 
Again,  $T_m,\v$ relax and $\sigma^e$ grows, though only until the arrow  cuts the critical 
$K\varepsilon^c$-line, at $r_T=0$. The third arrow has  $\sigma>K\varepsilon^e>K\varepsilon^c$, $r_T<0$, hence $T_m,
\v$ grow, while $\sigma^e$ decreases, until the critical state is 
reached. 
In the right figure, with $1/\tau=\lambda_{0.5}\sqrt{T_m}$, there is a critical state for any $\sigma^e$-value. And it is 
impossible to arrive at $\v=0$ without cutting the   $K\varepsilon^c$-line first.}
\end{center}
\end{figure}

For given stresses,  TYF have two steady-states to relax to. With $1/\tau=\lambda_1T_m$ and $\varepsilon^c=1/\lambda_1\ll1$, they are  
the critical state, $\sigma=K/\lambda_1 +\eta \v$,  $T_m=\v\not=0$, and stressed equilibrium,  $\sigma=\sigma^e<\sigma^{\rm Y}$ with  $T_m=\v=0$. This has been 
observed as viscosity bifurcation~\cite{vicBifurc,vicBifurc2}.  
The dynamics approaching these two states is still governed by 
Eqs.(\ref{eq3a}, \ref{eq8}).  
Inserting $\sigma=K\varepsilon^e+\eta\v=$ const into Eq.(\ref{eq3a}) to eliminate $\v$,  denoting $1/{\tau^\star}=1/\tau+K/\eta$, yields
\begin{align}\label{eq49}
\frac{\partial}{\partial t}\varepsilon^e=\v-\frac{\varepsilon^e}{\tau}=\frac{\sigma}{\eta}-\frac{\varepsilon^e}{\tau^\star}.
\end{align}
Assuming Eq.(\ref{eq1a}), or $\eta/K\ll1/r_m$, this is the {fast dynamics}, such that when solving Eq.(\ref{eq8}), the slow one, we may set the rhs.\,to zero, implying 
\begin{align}\label{eq27}
{\v}/{T_m}=\varepsilon^e\lambda_1=\varepsilon^e/\varepsilon^c.
\end{align}
Inserting this into  Eq.(\ref{eq8}) yields an equation central to stress-controlled dynamics, 
\begin{align}\label{eq51}
\z{T_m}=-r_TT_m,\quad r_T\equiv \textstyle\frac12 r_m\left[1-(\varepsilon^e/{\varepsilon^c})^2\right].
\end{align}
Clearly, $T_m$ relaxes until either $T_m=0$ or $r_T=0$, ie.~equilibrium or the critical state. [With Eq.(\ref{eq27}), $\varepsilon^e=\varepsilon^c$ implies $\v=T_m$.] 
For $\varepsilon^e>{\varepsilon^c}$, $r_T<0$, we have $T_m$ and $\v$ increasing, while $
\varepsilon^e$ decreases (to keep $\sigma=$ const), until they arrive at the critical state. For $\varepsilon^e<{\varepsilon^c}$, $r_T>0$, there are two cases: the fluid relaxes into 
equilibrium for $\sigma<K\varepsilon^c$, and becomes  critical for $\sigma>K\varepsilon^c$, see Fig.\,\ref{fig5}. 

The square bracket of Eq.(\ref{eq51}) is nearly const for $\eta/K\ll\tau$, because with 
$\z\sigma=K\z\varepsilon^e+\eta\z\v=0$,  we have $-\tau\z\v=(K\tau/
\eta)\z\varepsilon^e\gg\z\varepsilon^e$. This means $T_m$  evolves fairly exponentially, with a slowly vanishing rate $r_T$.  In Sec.\,\ref{sj}, we discuss why the rate-curve is monotonic in Fig.\,2 of~\cite{vicBifurc2}, but not in Fig.\,3.  

A  problematic implication of  Eq.(\ref{eq51}) is the fact that 
supercritical static states, $\sigma^{\rm Y}>\sigma^e>K\varepsilon^c$, are precarious, as infinitesimal ambient noises and $T_m$-fluctuations are predicted to grow exponentially,  
pushing the fluid into the critical state. On the other hand,  such supercritical  states are routinely observed---such as on a tilted plane, cf.\,Sec.\,\ref{tiltedP}.  
Therefore, the hybrid $\tau$ of Eq.(\ref{eq11}) may be closer to what is actually realized in nature as a yield-stress fluid.  
In this case, small $T_m$ will relax back to zero: Assuming $T_m$ is sufficiently small that $1/\tau=\lambda_2T^2$ holds, Eqs.(\ref{eq27},\ref{eq51}) become $\v=\varepsilon^e\lambda_2T_m^2$ and 
\begin{align}\label{stabil}
\z T_m^2=-r_m[T_m^2-(\varepsilon^e\lambda_2T_m^2)^2],
\end{align}
where the second term  $\sim T_m^4$, of higher order, may be neglected, leaving a simple relaxation.  
Only  sufficiently large $T_m$ will grow, destabilizing ab over-stressed, static state. You have to hit the ketchup bottle hard enough for the content to start flowing.

Finally,  we vary the parameterization of the relaxation time, as $1/\tau=\lambda_nT_m^n$. Because of the constraint, $
\v\tau\ll1$, we cannot chose  $n>1$, as $\v\tau$ diverges for $\v\to0$. Hence we consider $n=0.5$, or $\varepsilon^c(0.5)=\v\tau=(\v/|\v|)\sqrt{|\v|}/\lambda_{0.5}\ll1$, showing
that $n=1$ is a preferred choice: Only it displays viscosity bifurcation. 
With $\v=\varepsilon^e\lambda_{0.5}
\sqrt{T_m}$, we have 
${\v}/{T_m}=({\varepsilon^e}/{\varepsilon^c})^2$ that, 
inserted into Eq.(\ref{eq8}), yields 
\begin{align}\label{eq52}
\z{T_m}=-r_TT_m, \quad r_T=(r_m/2)\left[1-(\varepsilon^e/\varepsilon^c)^4\right].
\end{align}
The main difference to the previous case is that the critical state,  $\sigma=K\varepsilon^c(0.5) +\eta\v,\,\, \varepsilon^c(0.5)=\sqrt{\v}/\lambda_{0.5}$ exists for any $\sigma$-value. And there is no way for $\sigma^e$ to reach the $\sigma^e$-axis, with $T_m,\v=0$, from any point, see the left of Fig.\ref{fig5}. 
[The same holds for the hybrid $\tau$ of Eq.(\ref{eq11}). However, it may cross the the green line of Fig.\,\ref{fig5} at such low $\v$-values that is hard to distinguish from a full stop.]

\subsection{Stress Jumps\label{sj}}

After a stress jump, the rate change is known to be discontinuous for TYF and continuous for polymers, see Fig.\,1 of~\cite{SJ1}. This is easy to understand from the fast and slow dynamics of 
Sec.\,\ref{vibi}.   Discontinuity is of course a relative concept that depends on the timescale of the 
experiment. It appears discontinuous on the timescale of the slow dynamics,  Eq.(\ref{eq51}), and 
gets resolved by that of the fast dynamics, Eq.(\ref{eq49})---unless the time axis is logarithmic, then both stages are continuous.  

At the jump $\sigma_i\to\sigma_f$, it is $\varepsilon^e$ that changes instantly, to acquire the new stress value, 
\begin{align}\label{eqq37}
\sigma_i=K\varepsilon^e_i+\eta\v_i\longrightarrow\sigma_f=K\varepsilon^e_0+
\eta\v_i, 
\end{align}
The subscripts $i, 0,  2,  f$ stand, respectively, for \textit{pre-jump initial, post-jump initial,} 
\textit{intermediate} between the fast and slow 
 dynamics, and the \textit{final}  values. Assuming a critical state pre-jump, we have $\varepsilon^e_i=\varepsilon^c$ and $T_m=\v_i$. The fast relaxation, Eq.(\ref{eq49}), follows, creating, as mentioned above, a rate-discontinuity of $\v_2-\v_i$.  It ends with the values
\begin{align}\nonumber
\sigma_f&=K\varepsilon^e_2+\eta\v_2,\\\label{eqq39}
\varepsilon^e_2/\varepsilon^c&=\v_2/\v_i=\sigma_f/\sigma_i.
\end{align}
[The second line is given by taking $\v\to\v_2$, $T_m\to\v_i$ in Eq.(\ref{eq27}), 
and by writing $\sigma_i=K\varepsilon^c+\eta\v_i=(K\varepsilon^2_2+\eta\v_2)\v_i/\v_2=\sigma_f\v_i/\v_2$.] 
The ensuing slow relaxation,  Eq.(\ref{eq51}),  have the system arrive at the critical state, $\sigma_f=K/\lambda_1+\eta\v_f$, or a rest state. 

Next, we consider why the logarithmic rate-curve of~\cite{vicBifurc2} is monotonic in Fig.\,2, but not in Fig.\,3, after a 20~s  rest. (See also Fig.6 of~\cite{Larson},  in color.)
The plots of Fig.\,2 are stress jumps as considered above. They start with a pre-shear stress of 26 Pa,  jumping down to various stresses, with the highest $\sigma_f$ being 13.1 Pa. 
Eq.\,(\ref{eqq37}) states $\sigma_f<\sigma_i$ implies $\varepsilon^e<\varepsilon^c=1/\lambda_1$. Eq.(\ref{eq49}) has $\z\varepsilon^e>0$, for $\varepsilon^e/\varepsilon^c<\v/T_m$, as is the case initially, because $\v=T_m=\v_i$. Hence we have $\z\v=-(K/\eta)\z\varepsilon^e<0$, with no possibility for a sign change, as it is a relaxation equation. This goes on until  Eq.\,(\ref{eqq39}) holds, when the slow dynamics takes over. 
The slope of the slow dynamics depends on 
the actual value of the critical stress, $K \varepsilon^c$, that we may presume is close to 20 Pa from Fig.\,3, as the curve at 12 Pa clearly comes to rest, and that at 20 Pa goes critical,  probably. (Fig.\,5 has similar values.) Since the highest stress is 13.1 Pa, we have $ \varepsilon^e< \varepsilon^c$, implying $\z\v<0$,  see Eq.\,\ref{eq51}, such that all curves of Fig.\,2 are monotonic. The same  arguments have the curves remaining monotonic if the jumps were up.

For the data of Fig.\,3, the above chain of events is broken by a stress pause of 20 s (about the timescale of the fast dynamics), probably such that the pre-shear stress $\sigma_i$ comes down 
significantly, as does $\v_i$, though $T_m$ needs longer to relax (more so if the aging effect is operative, see Sec.\,\ref{aging}). Applying a stress of $\sigma_f$ after the pause creates 
mainly ${\varepsilon^e}$, see Eq.\,(\ref{eqq37}), also some $T_m$ (as it were not adiabatically applied), leaving $\v$ unchanged and small. This may be the reason ${\varepsilon^e}/{\varepsilon^c}>\v/T_m$ holds, implying (contrary to above) a positive rate-slope. The ensuing slow dynamics is not changed, with the curve at 12 Pa turning negative, and the one at 20 Pa remaining positive. 
(The 150~s of Fig.\,4 could have diminished $T_m$ as well, rendering the detailed process of applying the stress more relevant.) 

To solve   Eqs.\,(\ref{eq49},\ref{eq51}) explicitly, one needs to replace $\varepsilon^e$ in (\ref{eq51}) with $T_m$, as
$\varepsilon^e=\sigma/(K+\eta\lambda_1T_m)$.  It is left undone because this a complex 
experiment, and the result would only be convincing if one is confined by prior knowledge from simpler experiments, of the parameters $K,\eta,r_m$ and $1/\lambda_1$. Under the present circumstance,  what we have gained is a possible  explanation  of {viscosity bifurcation} that is qualitative, but unforced, coherent, and a direct result of Eqs.(\ref{eq3a},\ref{eq8})---a model obtained by simply interpolating   fluid- and solid-dynamics, not one sculpted to fit an experiment.  

For polymers, $\tau=$ const, there is only the fast dynamics, Eq.(\ref{eq49}). 
After a stress jump, $\sigma_i\to\sigma_f$, there is the same instantaneous change, $\varepsilon^e_i,\v_i\to\varepsilon^e_0,\v_i$, with an ensuing relaxation into the critical state, $\tau\v_f=\varepsilon^e_f$. On the timescale of $\tau^\star$,  the rate evolution is  continuous.

\subsection{Tilted Plane\label{tiltedP}}
\begin{figure}[t]
\begin{center}
\includegraphics[scale=.24]{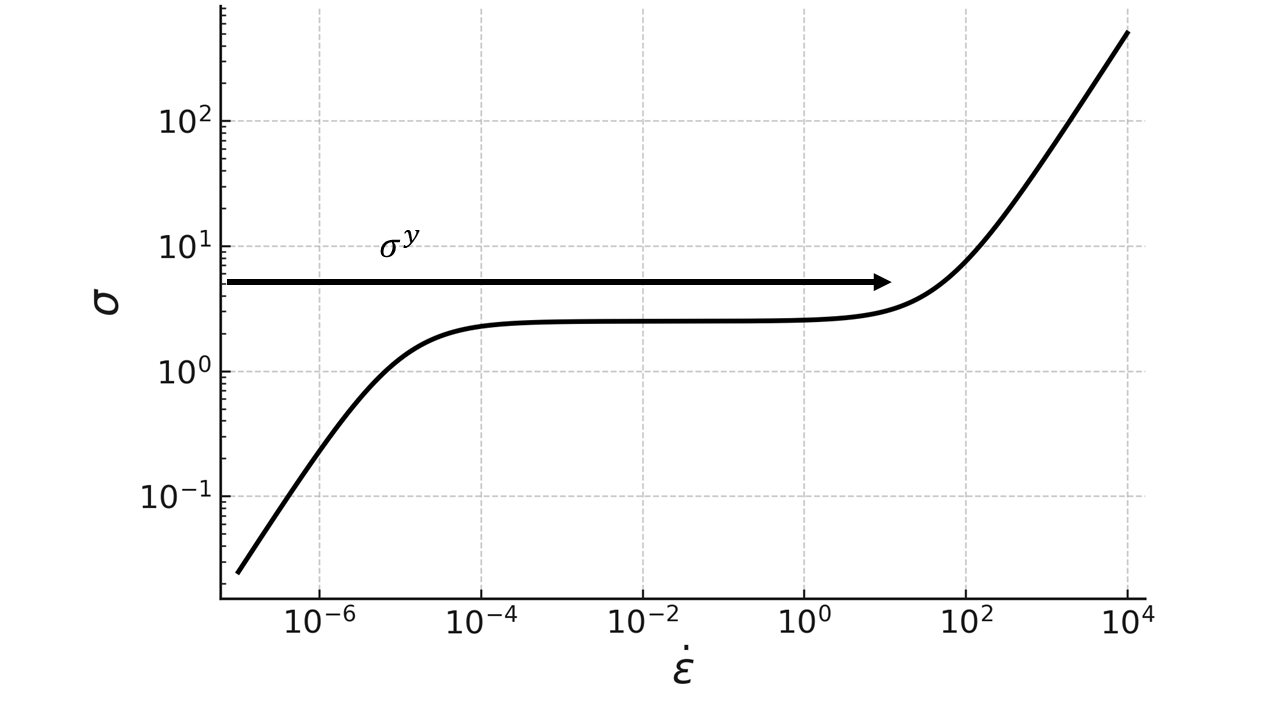}
\caption{\label{fig6} With sufficient tilt, the stress will breach $\sigma^{\rm Y}$, hence $\v$ grows, as prescribed by Eq.(\ref{eq51}), from $\v=0$  to where the stress cuts the critical line. (In contrast to Fig.\ref{fig5}), the y-axis is in $\sigma$, not $\sigma^e$, hence there is no tilt of the arrow.)}
\end{center}
\end{figure}
The tilted plane is a stress-controlled  set-up 
well suited for verifying some results from the last two sections.
Putting yield-stress fluid on a slowly tilting plane, the fluid will 
start flowing when the shear stress breaches the static yield-stress $\sigma^{\rm Y}$ (\textit{angle of stability}).  Assuming $1/\tau=\lambda_1T_m$ and a long horizontal stretch of the critical elastic stress, as in Fig.\ref{fig1}, the strain rate has to 
accelerate from zero, in accordance to Eq.(\ref{eq51}),  to cut the critical line, see Fig \ref{fig6}. And if the tilted plane is not long enough, 
one only observes the rate acceleration.
Reversing the tilt, the flow will slow down, until it stops,  (\textit{angle of  repose}), $\v=0$,  at the critical state, $\sigma^c=\sigma(\varepsilon^c)$, again as given by Eq.(\ref{eq51}), see~\cite{vicBifurc,vicBifurc2}. 



\section{Non-Uniform Experiments\label{nue}} 

Same as solid- or fluid-dynamics, TE is a model capable of dealing with spatial variations. Until now, we considered uniform $T_m,\varepsilon^e,\v$. In this section, we first consider two coexisting uniform systems, then continuous spatial variations. 

\subsection{Shear Band\label{sb}} 
\begin{figure}[t]
\begin{center}\vspace{-3cm}
\includegraphics[scale=.35]{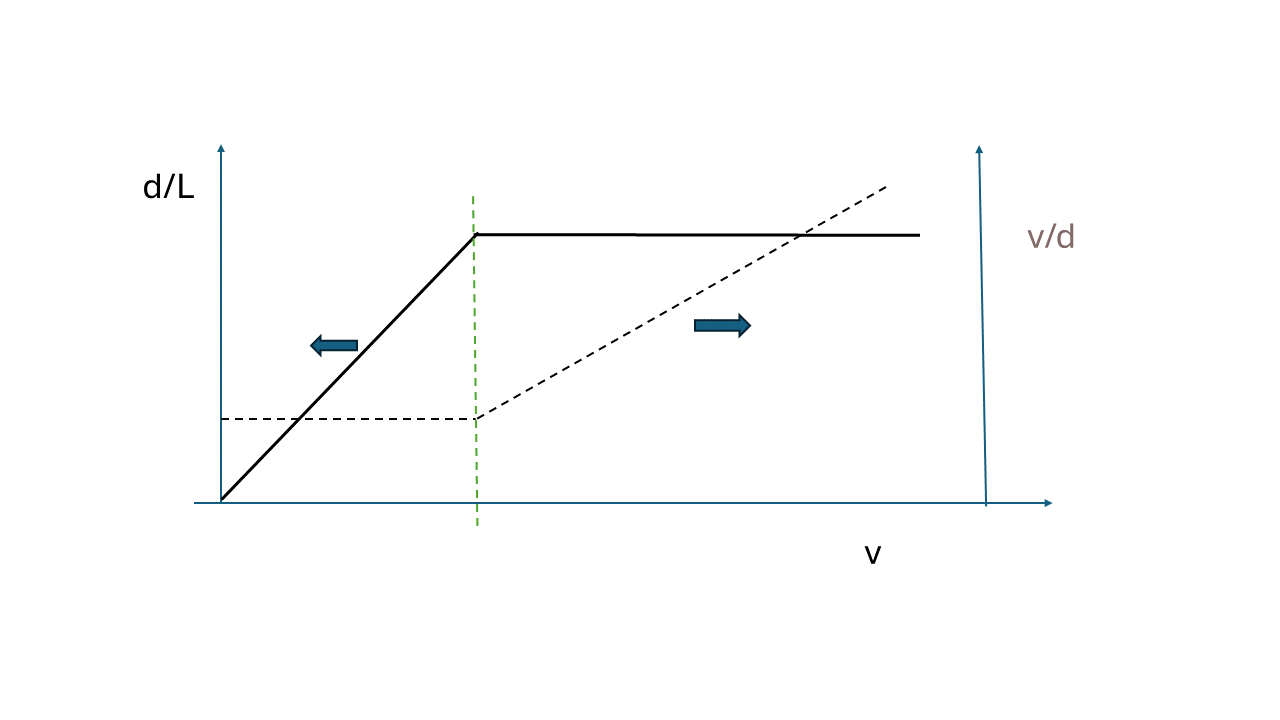}
\vspace{-1.8cm}
\caption{\label{fig7}With $L$  the system width, $d$ the shear band width, and $v$ the velocity difference, the straight black line depicts $d/L$ vs $v$. According to  Eqs.(\ref{eq26b}), $d\sim v$ as long as there is a shear band, until $d=L$. The broken black line depicts the local rate $\v=v/d$ that remains constant as $d$ grows. After $d/L=1$ is reached, the normal behavior of  $\v$ growing with $v$ is restored. These are as  reported in~\cite{SB1} and depicted in their Fig.\,2, which is quite the same as this plot.}
\end{center}
\end{figure}

A steady-state shear band may be taken as two steady-states in coexistence, one solid and quiescent, 
$\sigma_S=K\varepsilon^e$, the other fluid and critical, $\sigma_F=K\varepsilon^c+
\eta\v=K/\lambda_1+\eta\v$ (assuming $1/\tau=\lambda_1T_m$). Consider a 
1D-system, width $L$,  the solid of width $(L-d)$,  the fluid of $d$. The outer rim of the fluid moves at $v$, implying $T_m=\v=v/d$. 
(In truth, $T_m$ will spill slightly into the solid, see Sec.\,\ref{nf}.) The connecting conditions at the fluid-solid interface are
\begin{align}\label{eq26}
\sigma_S=\sigma_F\equiv\sigma= (K\varepsilon^e)_F/\gamma,
\end{align}
The first equal sign is the result of momentum conservation: 
Integrating $\z(\rho v_i)-\nabla_i\sigma=0$ over the volume of an an infinitesimally narrow slab containing the interface, and applying the Gauss Divergence Theorem, 
we obtain $\int\z(\rho v_i)\text{dVol}=\oint\sigma\text{dA}$, where the left side vanishes, implying $\sigma_F-\sigma_S=0$.
The $\equiv$-sign is a definition, and the second equal sign is an Onsager relation, with $\gamma=$ const. It is a result of considering the interface entropy production, see App.\,\ref{cc}. As the elastic stress of the fluid is constant, $(K\varepsilon^e)_F=K/\lambda_1$, so is the stress $\sigma$.  Hence the three  unknowns of the problem are connected as 
\begin{align}\label{eq26b}
\varepsilon^e_S=\frac\sigma K,\quad \v=\frac vd=\frac{\sigma-K/\lambda_1}\eta.
\end{align}
And we may conclude: $\varepsilon^e_S$, $\sigma$ and $\v$ are all constant, while $d$ grows with $v$, as long as there is a shear band. (Without it, $\v$ and $\sigma=K/\lambda_1+\eta(v/L)$ grow with $v$, as usual.) The relation $d\sim v$ was observed in~\cite{SB1} and depicted in Fig.\,\ref{fig7}.  

That $\sigma=$ const as long as there is a shear band,  independent of the velocity $v$, is a striking result, which apparently has not been reported. It is a direct consequence of the connecting condition $\gamma\sigma=\sigma^c_F$, derived in  App.\,\ref{cc} under a few assumptions that appear obvious but may not be generally true. Hence it is useful to be verified, or falsified, experimentally.

Unfortunately, circumstances are more complicated, as 
$\sigma$ must be smaller than $K\varepsilon^c$ in the solid  to be stable, cf.\,Sec.\,\ref{vibi}, but larger than $K\varepsilon^c$ in the fluid the system to flow, or 
\begin{align}\label{EQ37}
(K\varepsilon^c)_S>(K\varepsilon^e)_S=\sigma_S=\sigma_F=(K\varepsilon^c)_F+\eta\v>(K\varepsilon^c)_F.
\end{align} 
This we shall consider in more detail, while noting Eqs.(\ref{eq26b}) remain valid. As $K\varepsilon^c$ is typically a function of densities,  the inequality 
$(K\varepsilon^c)_S>(K\varepsilon^c)_F$ may imply a density  difference. 
TYF are all multi-component systems, we consider two: $\rho_1$ of the background fluid,  and $\rho_2$ of the substance forming the meso-structure that  causes elasticity and maintains $T_m$. Hence 
$K, \varepsilon^c$ depend on  $\rho_2$. We take $\rho_1$ as much less compressible,  varying only slightly to compensate the  $\rho_2$-discontinuity at the boundary, maintaining a uniform pressure. 
Both $K$ and  $\varepsilon^c=1/\lambda_1$ should increase with $\rho_2$, as a larger $\rho_2$ means more elasticity and longer $\tau=1/(\lambda_1T_m)$. If so, $\rho_2^S>\rho_2^F$ implies $(K\varepsilon^c)_S>(K\varepsilon^c)_F$, satisfying Eq.(\ref{EQ37}). 
Defining $\rho_2^F\equiv\rho_2-\delta\rho^F_2$, $\rho_2^S\equiv\rho_2+\delta\rho^S_2$ means  $\delta\rho^F_2, \delta\rho^S_2>0$. Given the mass, $M_2=\rho_2^Fd+\rho_2^S(L-d)$,  we have $
\delta\rho^F_2=\delta\rho^S_2(L/d-1)$, leaving only $\delta\rho^F_2$ as an additional unknown. It is determined by the condition of equal chemical potentials, expressing zero mass transfer across the 
interface. With the energy $w= \textstyle\frac12s_m^2/A+\textstyle\frac12K(\varepsilon^e)^2+w_0(\rho_1,\rho_2)$ and  $\rq{}\equiv\frac\partial{\partial\rho_2}$, the condition 
\begin{align}\label{eq26a}
&\mu^S_2=\mu^F_2\equiv\partial w/\partial\rho_2\quad\text{may be written as}
\\\nonumber
[(&\varepsilon^e)^2K\rq{}+w_0\rq{}]_{\rho^S_2}=[(\varepsilon^c)^2K\rq{}
-T_m^2A\rq{}+w_0\rq{}]_{\rho^F_2}.
\end{align}
This condition may be solved analytically by expanding around $\rho_2$, to linear order of $\delta\rho^F_2$. 
More generally, we take $\varepsilon^c=X_n/(4+X_n^2)$, see Eqs.(\ref{eq10}).  As $\sigma=K\varepsilon^c_F+\eta\v=K\varepsilon^c_F/\gamma$ or $\eta\v=K(1/\gamma-1)\varepsilon^c$, and   $X_n$ is a function of $\v$ alone, this expression determines $\v$ to be a constant,  same as  $\sigma$. In all these cases, we have $v\sim d$.

\subsection{Nonlocal Fluidization\label{nf}}
In this section, we complete the shear band picture of the last section, showing how $T_m$ spills into the solid portion, with decreasing $\v$, and leading to interesting  consequences. 

In Sec.\,\ref{sb}, we considered the solid-fluid coexistence, in which  the stress $\sigma$ is uniform, while $\rho_2$ and $T_m\sim\v$ are discontinuous. Yet  Eq.\,(\ref{eq8a}) prescribes a finite decay length for $T_m$. To ameliorate this, we note that since $\varepsilon^e$ is stationary,  Eq.\,(\ref{eq27}) holds (for both rate- and stress-control).  Inserting it into (\ref{eq8a}), we find $T_m,\v$ spatially decaying on the solid side, $x\ge0$, from the fluid temperature~$T_m^F$,
\begin{align}
\label{cr1}
\nabla^2_iT_m= \frac{T_m}{\xi^2_m},&\quad \xi_m^2\equiv\frac\kappa\beta\frac1{[1-(\varepsilon^e/\varepsilon^c)^2]},\\\nonumber
(\varepsilon^c/\varepsilon^e)\v= T_m&=T_m^F\exp(-x/\xi_m).
\end{align}
The velocity difference from this $T_m$-diffusion is $\int\v{\rm d}
x=\v_0\int_0^\infty \exp(-x/\xi_m){\rm d}x= \v_0\xi_m$, where $\v_0\equiv T_m^F(\varepsilon^e/\varepsilon^c)$. This implies $\v=(v-\v_0\xi_m)/d$ in the fluid, or 
\begin{align}\label{eq43}
T_m=\v=v/(d+\xi_m\varepsilon^e/\varepsilon^c).
\end{align}
[If $\beta\not\approx0$ in Eqs.(\ref{eq55}), $\Delta v=\beta\sigma$ needs to be added.] Assuming Eq.\,(\ref{eq1a}), the change of $\varepsilon^e$ (in the solid) due to the finite 
$\v$ is negligible. (The solid dynamics in fact has terms $\sim\nabla^2\varepsilon^e$, cf.\,Eq.\,(\ref{eqA4}) and~\cite{MPP}. The associated length scale is typically small and has been set to zero here, allowing a discontinuous change of $\varepsilon^e$.) So the modified $T_m, \v$-behavior  on the solid side are the only changes to the shear band results of Sec.\,\ref{sb}. As the decay length $\xi_m$ diverges for $\varepsilon^e\to\varepsilon^c$, there is then no solid left.

With $\tau=$ const, polymers, being  always in the fluid state for given shear stress, cannot suspend a ball with a higher density. In contrast, TYF can stay in 
the solid state, $T_m=0$, and suspend this ball (assuming it is not too heavy, such that $\sigma<\sigma^{\rm Y}$). However, $T_m$ spilled over from a shear band will fluidize the solid, because $1/\tau\not=0$ and the stress will relax. Under such \textit{non-local fluidization}, static stresses decay, and the yield-stress vanishes. A ball will sink,  or rise, depending on its density---as the Archimedes law prescribes.  (Periodic tapping has the same effects.) Same considerations holds for granular media, cf.\,\cite{exp11}.

\subsection{Narrow Shear Bands\label{nsb}}
The $T_m$-diffusion of  Eq.(\ref{cr1}) has a spatially confined solution for $\varepsilon^e>\varepsilon^c$, 
\begin{align}
\label{cr2}
\nabla^2_iT_m&= -T_m/\hat\xi^2_m,\,\, T_m=A\sin( x/\hat\xi_m),
\end{align}
where $A$ is the amplitude; and $\hat\xi_m^2\equiv(\kappa/\beta)/[(\varepsilon^e/\varepsilon^c)^2-1]>0$. It may be seen as two solid blocks slipping past each other, sandwiching a fluid layer of thickness $\pi \hat\xi_m$. Assuming strong elasticity, Eq.(\ref{eq1a}),  $
\sigma=$ const implies $\varepsilon^e\approx$ const, given by its solid value,  irrespective how $\v$  varies. The density $\rho_2$ is lower in the fluid (cf.\,Sec.\,\ref{sb}), such that $\varepsilon^c|_S>\varepsilon^e|_S>\varepsilon^c|_F$. 
Employing   Eq.\,(\ref{eq27}), or $\v= T_m(\varepsilon^e/\varepsilon^c)$, to calculate $A$, we have $v=\int\v{\rm d}x=A(\varepsilon^e/\varepsilon^c)\int_0^{\pi\hat\xi_m}\sin(x/\hat\xi_m){\rm d}x = 2A(\varepsilon^e/\varepsilon^c)\hat\xi_m$, or $A=(\varepsilon^c/\varepsilon^e)v/2\hat\xi_m$. 


\subsection{Elastic Shear Waves\label{waves}}
Employing TE and assuming $1/\tau=\lambda_1T_m$, we consider the propagation of elastic shear waves in TYF, showing that they should exist for high frequencies, $\omega\tau\gg 1$, even post-yield, as long as the wave lengths is large compared to meso-structures.

We start from  $\rho\z v=K\nabla\varepsilon^e$, 
$\z\varepsilon^e=\v-\varepsilon^e/\tau$, see Eqs.(\ref{eq1},\ref{eq2}),
with $v=v_1$, $\nabla=\nabla_2$. (The terms $\rho v_iv_j$ and $\eta\v _{ij}$ are neglected, because  the first is quadratic in $v_i$, and $\varepsilon^e_{ij}/\tau$ delivers the typically dominant damping, larger than the viscous stress.)
Together, they form the  telegraph equation, 
\begin{align}
2\rho\left(\frac{\partial^2}{\partial t^2}+\frac1\tau\frac\partial{\partial t}\right)\varepsilon^e=K\nabla^2\varepsilon^e.
\end{align}
With $\delta\varepsilon^e\sim e^{iqx-i\omega t}$, we find
$c^2 q^2={\omega^2+i\omega/\tau}$ with $c^2\equiv K/2\rho$, or
\begin{align}\label{eq62}
cq=\pm\omega \left(1+i/{2\omega\tau}\right),
\end{align}
propagating for $\omega\tau\gg1$. 
We take  $1/\tau=\lambda_1T_m$ and $\omega\gg r_m$, such that Eq.(\ref{eq61}) holds,with $T_m\approx|\v_0|=$ const. Inserting  Eq.(\ref{eq62}) into  $\delta\varepsilon^e\sim e^{iqx-i\omega t}$, we find 
\begin{align}
\delta\varepsilon^e\propto e^{\left[-i\omega\left(t\mp x/c \right) \mp x({\lambda_1 T_m}/{2 c})\right]},
\end{align}
or that the shear wave  propagates for the distance of approximately $2c/\lambda_1T_m$. Placing a piezoelectric detector around this distance, one should be able to pick up the wave signal. The fluid to choose should have strong elasticity, with $K$ large; the steady shear $\v_0$ should be kept small, as should the wave amplitude $\delta\v$. 

Without a steady shear rate, the elastic shear wave will always propagate for the initial time span $t\le 1/r_m$, because $T_m\approx0$ then, and for all time $t$ if the system is accounted for by the hybrid relaxation time of Eq.(\ref{eq11}).

\section{Conclusion\label{1-4}}
The prevalent picture of thixotropic yield stress fluids is that of a fluid with a complex viscosity.  This paper suggests that it may simply be taken as \textit{``not quite solid''}: Interpolating solid- and fluid-dynamics and incorporating a second temperature for the mesoscopic structure, yields the model called TE, for transient elasticity. It blurs the  line between fluid and solid,  accounting for a system that flows readily, appears liquid, yet manifests its elastic, solid part by producing effects at  rate and stress jumps that we call non-Newtonian, including overshoots, viscosity bifurcation, shear band, shear waves, and oscillatory rheography. 
Some of the more interesting results are summarized below:

\textbf{Rate jumps} and \textbf{rate ramps} are typical rate-controlled experiments, accounted for by 
Eq.(\ref{funda}) directly. For polymers, one has $\tau=$ const. With only one timescale, the stress relaxation after a rate jump is  monotonic. 
For granular media and TYF, there are two timescales, $\tau\sim 1/T_m$ for elastic relaxation and $1/r_m=$ const for the equilibration $T_m\to0$, rendering  non-monotonic stress relaxation possible, eg.\,overshoots. As the first timescale depends on the second, its relaxation behavior is  intricate. 
In granular media, the relaxation $T_m\to\v$ is fast, typically $1/r_m\ll\tau$, such that right after the rate jump $\v_i\to\v_f$, elasticity starts to relax with $\tau\sim1/\v_f$. Hence one is back at the one-timescale case, and overshoot is not routinely reported. 
In the opposite case, $1/r_m\gg\tau$, the stress curve is discontinuous at $t=0$, as explained in  Sec.\,\ref{overshoot}. This case is drawn in Fig.1(c) of both~\cite{Larson,thix1} and labeled as \textit{``inelastic/ideal thixotropic''}. 
Taking both timescales as comparable, appropriate for TYF, leads to continuous, fully-developed over- and undershoots, as in Fig.\ref{fig4}.

\textbf{Oscillatory  rheography: }
The oscillatory stress rates, $G^\prime$ and  $G^{\prime\prime}$, linearly calculated, matches the observed {amplitude-dependence} well~\cite{aging}. This puzzling result is easy to understand for an  TE-system, less so for a viscous one: Given the Maxwell/TE-dynamics,  $G^\prime, G^{\prime\prime}$
are functions of $\omega\tau=\omega/\lambda_1T_m$. For $\omega\gg r_m$, instead of following $\v$ instantaneously, $T_m$ grows slowly  until $T_m=\langle\v\rangle=\omega\bar\gamma$, see Eq.(\ref{eq8}), or $\omega\tau=1/\lambda_1\bar\gamma$, where  $\bar\gamma$ is the oscillation amplitude. However, the frequency dependence of  $G^\prime, G^{\prime\prime}$ returns under a steady shear. 

\textbf{Viscosity bifurcation: }
For stress-controlled dynamics, one needs to insert $\sigma=$ const into Eq.(\ref{funda}) to eliminate $\v$. This changes the relaxation rate for  $\varepsilon^e$, from $1/\tau$ to $1/{\tau^\star}\equiv1/\tau+ K/\eta$, see Sec.\,\ref{vibi}. In critical state, the stress is $\sigma=(K\tau+\eta)\v$. As the first term is responsible for non-Newtonian effects, thixotropy is prominent, well-developed, if $\eta/K\ll\tau$. Taking $\tau$ and $1/r_m$ as comparable, we  proceed in Sec.\,\ref{sj} with  
\begin{align}\label{eq1a}
\eta/K\ll\tau, 1/r_m.
\end{align}
Hence $1/{\tau^\star}\approx K/\eta=$ const, and the dynamics of $\varepsilon^e$ becomes simple, as there are now two independent relaxations, a fast one for  $\varepsilon^e$ 
and a slow one for $T_m$. When solving the latter, we may take the former to have run its course, ${\v}/{T_m}=\varepsilon^e/\varepsilon^c$. This  
alters the relaxation rate of $T_m$, from $r_m$ to $r_T\equiv (r_m/2)\left[1-(\varepsilon^e/{\varepsilon^c})^2\right]$. As a result, the dynamics of  $T_m$ will stop either in equilibrium, for $T_m=\v(\varepsilon^c/\varepsilon^e)=0$, or at the critical state, $\varepsilon^e={\varepsilon^c}$, when the rate vanishes.  This is the TE-account of viscosity bifurcation. And since the fast and slow dynamics have different slopes, the logarithmic $\v(t)$-curve may well be  non-monotonic.

\textbf{Shear band: }
In TE, a shear band is taken as the coexistence of two states, 
critical and static, in Sec.\,\ref{sb}. The shear band width $d$ is shown to grow with the velocity difference $v$,  such that the local rate $\v=v/d=$ const~\cite{SB1}.  
Because $T_m$-diffuses into the solid part, the Archimedes law should hold in the vicinity of a shear band, see Sec.\,\ref{nf}. 

\textbf{Static yield: }
Because \textit{static yield} occurs in a quiescent system, at equilibrium,  it must be encoded in the energy.  Hence $\varepsilon^e=\varepsilon^{\rm Y}$ needs to be taken as the \textit{inflection point} of the  elastic energy $w(\varepsilon^e)$, 
where $w$ ceases to be  convex, rendering elasticity unstable for $\varepsilon^e>\varepsilon^{\rm Y}$. 
We have $\varepsilon^{\rm Y}>\varepsilon^{c}$ (the critical strain), because $\varepsilon^{c}$ is an elastic solution, while $\varepsilon^{\rm Y}$ is where elasticity ends. 
Because of thermodynamic stability, $\partial\sigma^e/\partial\varepsilon^e=\partial^2 w/(\partial\varepsilon^e)^2>0$, the elastic stress $\sigma^e(\varepsilon^e)$ is a monotonic function. Hence $\sigma^{\rm Y}>\sigma^{c}$ also holds. This is why the stress that initiates a flow, $\sigma^Y$, exceeds  the one sustaining it, $\sigma^c$~\cite{slibar}.  
Nether $\varepsilon^{\rm Y}$ nor  $\sigma^{\rm Y}$ are material constants, because elastic coefficients are functions of temperature and pressure. 

On a \textbf{tilted plane}, when the stress breaches $\sigma^{\rm Y}$, elasticity collapses, converting the elastic energy to $\v$ and $T_m$, and the fluid starts flowing. The initial state will converge gradually onto the critical state, see Sec.\,\ref{vibi}.  If it has a long horizontal stretch, as in Fig \ref{fig6}, the strain rate has to accelerate to cut the line on the far end. If the tilted plane is not long enough, one only observes the acceleration, as in~\cite{vicBifurc,vicBifurc2},  not the final steady-state. \\

\textit{Declaration of Interests: The author reports no conflict of interest.}

 \appendix
 \section{Hydrodynamic Theories\label{hydro}}
In condensed matter physics, the qualification \textit{hydrodynamic} goes well beyond Newtonian fluids. It  is employed to refer to any continuum mechanical  (or long wave length) model that 
complies with general principles of physics, including especially energy conservation and entropy production obeying the Onsager relation. In the main text, I use the term \textit{compliant} instead, 
because many rheologists deem the generalization of a term as  familiar as \textit{hydrodynamics} offensive and presumptuous. 

There are four groups of member systems, each with its established hydrodynamic theory: 
Newtonian fluids~\cite{LL6},  liquid crystals~\cite{deGennes,MPP,dGb} (including subgroups such as nematics, cholesterics and smectics), 
superfluids~\cite{LL6,Khal,Josephson2}, and solids~\cite{MPP,LL7,Lub2}.  Every member obeys its hydrodynamics, which are generally accepted as \textit{complete},  in the sense that all macro-scale experiments are accounted for, but no more---there are no phantom predictions. (All macro-scale  theories have the same known limits, such as the size of infinitesimal volume elements,  the time needed for local equilibrium to form,  also relativistic effects, see eg.\,\cite{acausal}.)  
 A prerequisite for this is \textit{compliance}:  Real systems in nature are compliant, they conserve energy and have a proper entropy production.  If a model fails at compliance, it is bound to disagree with some observations. Having constructed a non-compliant model for some experiments, it is doubtful whether other experiments will abide by the same model. 
 
One may be tempted to take these hydrodynamic theories as random constitutive models that its respective  members happens to comply with, implying the groups are uniform, with members that are similar. Yet this is hardly true: Solids can be metals, insulators, or semi-conductors that, 
irrespective of their atomic composition and interaction, 
all obey elasticity. Liquid helium (both $^4$He and $^3$He) is superfluid at low temperatures, same as cold atomic gases, and ultra dense neutron star material. Newtonian fluids and liquid crystals include materials of even more diverse atomic and chemical compositions. 

Generally speaking, given a set of state variables, the \textit{structure} of its hydrodynamic theory,  the form of its equations,   is fixed by adherence to general principles. Interactions and composition only 
influence how energy and transport coefficients depend on state variables, which are the \textit{constitutive choices}.  The structure differentiates the groups, the constitutive choices member systems. 
A hydrodynamic theory is a set of partial differential equations for its state variables, including mass,  entropy and momentum densities. They all have unique time-inversion parity: Energy, mass, entropy and the elastic strain are positive, momentum  odd. Their respective fluxes have dissipative and reactive terms, with parities that are the same or opposite to that of the state variable. Only the former contribute to entropy production.\,\cite{OnsagerLIT}  
The energy is a function of the state variables, while the entropy production $R$ depends on \textit{thermodynamic forces} such as $\nabla_iT$ or $\v\equiv\frac12(\nabla_iv_j+\nabla_jv_i)$. 
State variables characterize the equilibrium of a volume element; the forces measure how far equilibrium is, by comparing the state variables of neighboring volume elements. The Onsager relations guaranteeing the \textit{compliance} of a theory cover only the lowest order expansion of $R$ in the forces. In this order, the viscosity $\eta$ depends on state variables, not the forces. 
Going to an $\v$-dependent $\eta$ implies higher order terms while ignoring the generalized Onsager relations.

\section{Dynamics of Temperatures\label{dd}}
The evolution equations for the two temperatures are derived in Sec.\,(7.1) of~\cite{exp31}, though calling them $T$ and $T_g$. A brief summary is given below. We start with the energy $w$,
\begin{align}\nonumber
{\rm d}w=\mu{\rm d}\rho+T{\rm d}s+T_m{\rm d}s_m+v_i{\rm d}g_i+\sigma^e_{ij}{\rm d}\varepsilon^e_{ij},
\,\,\text{or}
\\\label{eqx1} \z w=\mu\z\rho+T\z s+T_m\z s_m+v_i\z g_i+\sigma^e_{ij}\z\varepsilon^e_{ij},
\end{align}
into which the following equations of motion,  expressing conservation laws and positive entropy production (where $R,R_m>0$), are inserted,  
\begin{align}\label{eqx2}
&\z w+\nabla_iQ_i=0,\quad \z g_i-\nabla_j\sigma_{ij}=0,
\\\nonumber &\z\rho+\nabla_iJ_i=0,\,\,  D_t\varepsilon^{e}_{ij}=\dot \varepsilon_{ij} -  \varepsilon^{e}_{ij}/{\tau}-\textstyle{\frac12}(\nabla_iy_j+\nabla_jy_i)
\\\nonumber
&\z s+\nabla_iF_i=R/T,\quad \z s_m+\nabla_iF^m_i=-\beta T_m+R_m/T_m,
\end{align}
\begin{align}
\text{yielding}\quad 
&\nabla_iQ_i=\mu\nabla_iJ_i-v_i\nabla_j\sigma_{ij}+T\nabla_iF_i+T_m\nabla_iF^m_i\\\nonumber&+\sigma^e_{ij}  [\dot \varepsilon_{ij} -  \varepsilon^{e}_{ij}/{\tau}-\textstyle{\frac12}(\nabla_iy_j+\nabla_jy_i)]-(R+R_m).
\end{align}
All terms following \lq\lq{}$\nabla_iQ_i=$\rq\rq{} can either be  rewritten as $\nabla$ of something and be part of $Q_i$, or included in $R+R_m$, noting that  $R,R_m=0$ have to hold in equilibrium. One result is
\begin{align}\label{eqA3}
Q_i=\mu J_i+TF_i+T_mF^m_i-v_k\sigma_{ik}-\sigma^e_{ik}y_k.
\end{align}
Further employing the Onsager relation (assuming isotropy) and requiring $R,R_m>0$ off equilibrium  suffice to determine  all fluxes above, see~\cite{exp31}, especially 
\begin{align}\label{eqA4}
y_i=\beta^p\nabla_k\sigma^e_{ik},\quad \beta^p= \text{const}.
\end{align}
This expression is useful for the \textit{connecting conditions} of Sec.\,(\ref{cc}). In Eqs.\,(\ref{eq2}),  we have $y_i\equiv0$, as we did not consider any non-uniform $\sigma^e_{ik}$, see the remarks below Eq.\,(\ref{eq43}). 

Here, we are interested in the equations for the entropy densities, $s$ and $s_m$, which are
\begin{align}
\label{2-17}
&\partial_ts+\nabla_i(sv_i-\kappa_1\nabla_iT)
=R/T,
\\\label{2-18a}
&R=\eta_1 \v^2+\kappa_1(\nabla_iT)^2+(K_1/\tau)(\varepsilon^e)^2+\beta T_m^2,
\\\label{2-19}
&\partial_ts_m+\nabla_i(s_mv_i-\kappa_2\nabla_i T_m)
=(R_m-\beta T_m^2)/ T_m,
\\ &R_m=\eta_2 \v^2+ \kappa_2(\nabla_iT_m)^2+(K_2/\tau)(\varepsilon^e)^2>0. \end{align}
Entropy production $R$ accounts generally for the transfers of energy into heat in both fluids and solids. In TE-systems, the first three terms account for a direct transfer of macroscopic into the microscopic  energy, only the last one transfers mesoscopic into microscopic energy. 
The meso-entropy $s_m$ has a similar structure, 
where  $R_m$ accounts for the dissipation to the first of the \textit{two-stage irreversibility}. 
As a constitutive choice, we take $\eta_1,K_2=0$: The first to simplify TE, and the second to have a stable 
stressed equilibrium below the yield stress.  [We also re-denote $\eta_2\to\eta, \kappa_2\to\kappa$, cf.\,Eq.(\ref{eq6}).] The term  $-\beta T_m^2$ accounts for the loss of meso-heat by $T_m$-relaxation. It has the same magnitude as the corresponding term in Eq.(\ref{2-18a}), such that no energy is lost.

\section{Connecting Conditions\label{cc}}  

Starting with Eqs.\,(\ref{eqA3},\ref{eqA4}) and taking  the normal unit vector of the  interface along $\hat y$, we denote the parallel fluxes of energy, mass 1, mass 2, entropy, meso-entropy, stress, respectively as $Q=Q_y$, $j_1=j^1_y$, $j_2=j^2_y$, $f=f_y$, $f_m=f^m_y$, $\sigma=\sigma_{xy}$,  $\sigma^e=\sigma^e_{xy}$, $v=v_x$, $y=y_x$. 
[We consider the two-densities case here: $Q_i=\mu^1 J_i^1+\mu^2 J_i^2+\cdots$, cf.\,Eq.(\ref{eqA3}).]
The energy flux traversing the interface is then
\begin{align}
\label{hy1S}
Q=\mu_1 j_1+\mu_2 j_2+Tf+T_mf_m-v\sigma-\sigma^ey,
\end{align}
see~\cite{alert}. Denoting $\Delta A\equiv A_S-A_F$ for any $A$ from the solid and fluid side, we have  $\Delta(AB)=A_SB_S-A_FB_F=\frac12(A_S+A_F)(B_S-B_F) +\frac12(B_S+B_F)(A_S-A_F)\equiv\bar A\Delta B+\bar B\Delta A$.    
Identifying the  interface entropy production as $R_{if}\equiv\bar T\Delta f+\bar T_m\Delta f_m$, we insert Eq.\,(\ref{hy1S}) into into $\Delta Q=0$ (an expression of energy conservation across the interface), while noting mass and momentum conservation, $\Delta j_1=\Delta j_2=\Delta\sigma_i=0$, to arrive at 
\begin{align}
R_{if}= j_i\nabla\mu_1+ j_2\nabla\mu_2+\bar f\nabla T+\bar f_m\nabla T_m+\sigma\Delta v+\sigma^e_Fy_F-\sigma^e_Sy_S,
\end{align}
Next, we take the first four terms to be diagonal, with $j_i\sim\nabla\mu_1$, $j_2\sim\nabla\mu_2$, $\bar f\sim\nabla T$, and $\bar f_m\sim\nabla T_m$, 
leaving a 3x3 Onsager matrix structure, 
\begin{align}
\label{DeGroot1}
\begin{pmatrix}
  \sigma^{e}_F\\  \sigma^{e}_S   \\  \Delta v_i
 \end{pmatrix}
=
\begin{pmatrix}
 \alpha_F & \alpha_x & \gamma_F\\
\alpha_x & \alpha_S & \gamma_S\\
-\gamma_F &-\gamma_S & \beta
 \end{pmatrix}
\!\times\!
\begin{pmatrix}
y^D_F\\-y^D_S\\\sigma
 \end{pmatrix}\!\!,
\end{align}
with $\alpha_F,\alpha_S,\beta, \alpha_F\alpha_S-\alpha_x^2>0$. For the circumstance of Sec.\,\ref{sb}, we have  $y^D_F,y^D_S= 0$ (for  uniform $\sigma^{e}$), see Eq.\,(\ref{eqA4}), and  $ \sigma^{e}_S=\sigma$. Hence $\gamma_S=1$, and the connecting conditions are   
\begin{align}\label{eq55}
\sigma^{e}_F=\gamma_F\sigma,\quad \Delta v=\beta\sigma. 
\end{align}
In Sec.\,\ref{sb}, we employ $\sigma^{e}_F=\gamma\sigma$, and assume $\beta\to0$, ie. $\Delta v=0$ to simplify the results.

\section{Glossary\label{Glossary}}
\noindent
\textbf{TYF:} Thixotropic Yield-Stress Fluid, typically with $1/\tau=\lambda_1T_m$. \\

\noindent
\textbf{State variables} are the variables of the energy density $w$, which in TE are the 
elastic strain $\varepsilon^e_{ij}$ and the densities of entropy $s$, meso-entropy $s_m$, various masses $\rho_\alpha$ (with $\rho=\sum\rho_\alpha$), and momentum $g_i=\rho v_i$.  
Their conjugate variables are respectively 
chemical potentials $\mu_\alpha=\partial w/\partial\rho_\alpha$, temperature $T=\partial w/\partial s$,  meso-temperature $T_m=\partial w/\partial s_m$, velocity $v_i=\partial w/\partial g_i$, and elastic stress $\sigma^e_{ij}=\partial w/\partial\varepsilon^e_{ij}$.\\

\noindent\textbf{Mesoscopic Temperature $\bm{T_m}$} accounts for the fluctuating kinetic and elastic energy of the meso-structures in TYF. In equilibrium, $T_m=T$ holds.     \\

\noindent
\textbf{Elastic Strain $\bm{\varepsilon^e_{ij}}$} is the reversible, energy-storing part of the strain, where $\varepsilon^e_{12}\equiv\varepsilon^e$. The critical value of the elastic strain, $\varepsilon^c_{12}\equiv\varepsilon^c$, is the stationary solution after both $\varepsilon^e$ and $T_m$ have fully relaxed.  
\\ 

\noindent
\textbf{Stress and Yield:} 
$\sigma_{12}\equiv\sigma=\sigma^e+\rho v_1v_2+\eta\v$, with  $\sigma^e=\partial w/\partial\varepsilon^e$. Mostly, we take $\sigma^e=K\varepsilon^e$ with $K$ the elastic coefficient. Static yield-stress $\sigma^{\rm Y}\equiv\sigma^e(\varepsilon^{\rm Y})$ is the inflection point $\varepsilon^{\rm Y}$ of  $w(\varepsilon^e)$. Dynamics yield $\sigma^{c}=\sigma^e(\varepsilon^c)$ is the stationary solution.\\

\noindent
\textbf{Rate  $\bm{\v}\equiv\varepsilon$:} The total rate is $\varepsilon\equiv\frac12(\nabla_1v_2+\nabla_2v_1)=D_t\varepsilon^e+\varepsilon^p$, with $\varepsilon^p=\varepsilon^e/\tau$.  The stress $\sigma$ depends on $\varepsilon^e$ and $\v$. Forcing it to depend on a fictitious total strain, $\varepsilon\equiv\int\v \textrm{d}t$, leads to hysteresis (cf.  Sec.\,\ref{inversion}).

\bibliographystyle{ieeetr}  
\bibliography{references.bib} 
\end{document}